# Driving Conditions-Driven Energy Management for Hybrid Electric Vehicles: A Review

Teng Liu[a,b], Wenhao Tan[b], Xiaolin Tang[b,*], Jinwei Zhang[a], Yang Xing[c], Dongpu Cao[a]

[a]Department of Mechanical and Mechatronics Engineering, University of Waterloo, Ontario N2L3G1, Canada;

[b]Department of Automotive Engineering, Chongqing University, Chongqing 400044, China;

[c]School of Mechanical and Aerospace Engineering, Nanyang Technological University, Singapore 639798;

*Corresponding author. Email address: tangxl0923@cqu.edu.cn

**Abstract**: Motivated by the concerns on transported fuel consumption and global air pollution, industrial engineers and academic researchers have made many efforts to construct more efficient and environment-friendly vehicles. Hybrid electric vehicles (HEVs) are the representative ones because they can satisfy the power demand by coordinating energy supplements among different energy storage devices. To achieve this goal, energy management approaches are crucial technology, and driving cycles are the critical influence factor. Therefore, this paper aims to summarize driving cycle-driven energy management strategies (EMSs) for HEVs. First, the definition and significance of driving cycles in the energy management field are clarified, and the recent literature in this research domain is reviewed and revisited. In addition, according to the known information of driving cycles, the EMSs are divided into three categories, and the relevant study directions, such as standard driving cycles, long-term driving cycle generation (LT-DCG) and short-term driving cycle prediction (ST-DCP) are illuminated and analyzed. Furthermore, the existing database of driving cycles in highway and urban aspects are displayed and discussed. Finally, this article also elaborates the future prospects of energy management technologies related to driving cycles. This paper focusing on helping the relevant researchers realize the state-of-the-art of HEVs' energy management field and also recognize its future development direction.

| | | | |
|---|---|---|---|
| **Nomenclature** | | HETV | Hybrid electric tracked vehicle |
| | | ICE | Internal combustion engine |
| | | ITS | Intelligent transportation system |
| A-ECMS | Adaptive ECMS | JC08 | Japan cycle'08 |
| ANN | Artificial neural network | $k$-NN | $K$-nearest neighbor |
| BA | Bee algorithm | LVQ | Learning vector quantization |
| BEVs | Battery electric vehicles | LT-DCG | Long-term driving cycle generation |
| CP | Convex programming | MC | Markov chain |
| DL | Deep learning | MPC | Model predictive control |
| DNN | Deep neural network | MTF | Mean tractive force |
| DP | Dynamic Programming | NEDC | New European driving cycle |
| DRL | Deep reinforcement learning | NGSIM | Next generation simulation |
| DQL | Deep Q-learning | NN | Neural network |
| DQN | Deep Q-network | NREL | National renewable energy laboratory |
| DDPG | Deep deterministic policy gradient | PHEVs | Plug-in hybrid electric vehicles |
| ECMS | Equivalent consumption minimization strategy | PMP | Pontryagin's minimum principle |
| | | PSO | Particle swarm optimization |
| EMs | Electric motors | PPO | Proximal policy optimization |
| EMS | Energy management strategy | RL | Reinforcement learning |
| ESC | Extremum seeking control | SA | Simulated annealing |
| ESSs | Energy storage systems | SDP | Stochastic dynamic programming |
| EUDC | Extra urban driving cycle | SMC | Sliding mode control |
| FTP | Federal test procedure | SOC | State of charge |
| FCHEV | Fuel cell hybrid electric vehicle | ST-DCP | Short-term driving cycle prediction |
| GA | Genetic algorithm | SOH | State of health |
| GIS | Geographical information systems | SVM | Support vector machine |
| GPS | Global positioning system | T-ECMS | Telemetric ECMS |
| GT | Game theory | TPMs | Transition probability matrices |
| HEVs | Hybrid electric vehicles | TSDC | Transportation secure data center |
| HWFET | Highway fuel economy test | UDC | Urban driving cycle |

## 1. Introduction

As we know, the consumption of fossil fuel is widely known as the primary cause of pollutant emissions and energy shortage. Increasing fuel-saving concerns and environmental awareness propel academic researchers and industrial engineers to search for more green and efficient solutions for the automotive industry [1, 2]. Vehicle electrification is regarded as a promising technology in last ten years to address this problem, and many manufacturers have produced different kinds of electrified vehicles, such as battery electric vehicles (BEVs), hybrid electric vehicles (HEVs) and plug-in hybrid electric vehicles (PHEVs) [3-7]. These vehicles contain more than one energy storage systems (ESSs), such as internal combustion engine (ICE), lithium-ion battery pack, fuel cell, and super-capacitor. Owing to this architecture, these kinds of vehicles could improve the overall powertrain efficiency by integrating electric motors (EMs) and special transmission device into the vehicle [8, 9]. Hence, hybrid vehicles currently dominate the sales market for electrified vehicles.

To take advantage of the HEVs' superiority, energy or power management is one of the most challenging problems for this complex system. It means distributing energy supplements for different ESSs while satisfying the physical constraints and predefined objectives [10, 11]. These objectives are represented as reducing exhaust emissions, delaying battery aging, maintaining vehicle mobility and drivability, minimizing fuel consumption, and so on. The constraints refer to the variables of onboard components, such as the state of charge (SOC) and state of health (SOH) in the battery, speed and torque limitations for the ICE and generator set, the output current and power for the battery pack, and dynamic characteristics of the powertrain [12, 13]. Furthermore, the most challenging problem in the energy management field of HEVs is that the driving conditions are always unknown to the vehicle control unit [14, 15]. The driving conditions here can be interpreted as the driving cycles, different driver styles and intention, road grade, outdoor temperature, and surrounding traffic situations. Without this information, the energy management controller cannot manipulate the power split reasonably among multiple energy resources, which results in a waste of fuel and energy.

Driving cycles discussed in this article indicate the vehicle velocity or speed trajectory, which is an indication of vehicle velocity/speed versus sample time [16]. It can capture the characteristics of acceleration, driving style, and driver behaviors [17, 18], and its main features are travel distance, duration time, average speed, and average acceleration. In HEVs' energy management problem, driving cycles are incredibly significant because they directly affect the power demand (since the powertrain parameters are specified), and thus they will influence the equilibrium of power flow between the ICE and other ESSs [19]. Moreover, the control performance of energy management strategy (EMS) depends heavily on the type of driving cycles, which had attracted a considerable amount of interests from researchers to derive an optimal EMS on a particular driving cycle [20]. Fig. 1 shows the relationship of the driving cycle, energy management controller, and hybrid powertrain. As the driving cycle is discerned as the input of the energy management problem, its information would affect the control performance of EMS extremely. However, driving cycles are partially or totally unknown in real-world driving situations, therefore, an online EMS should be mutable and adaptive. With that characteristic, this EMS could adapt to different driving cycles, and thus may be applied in online implementation. As a matter of fact, a few of current literature has put this thought in reality, and none of them has systematically summarized and categorized the existing techniques for driving cycle-related EMSs.

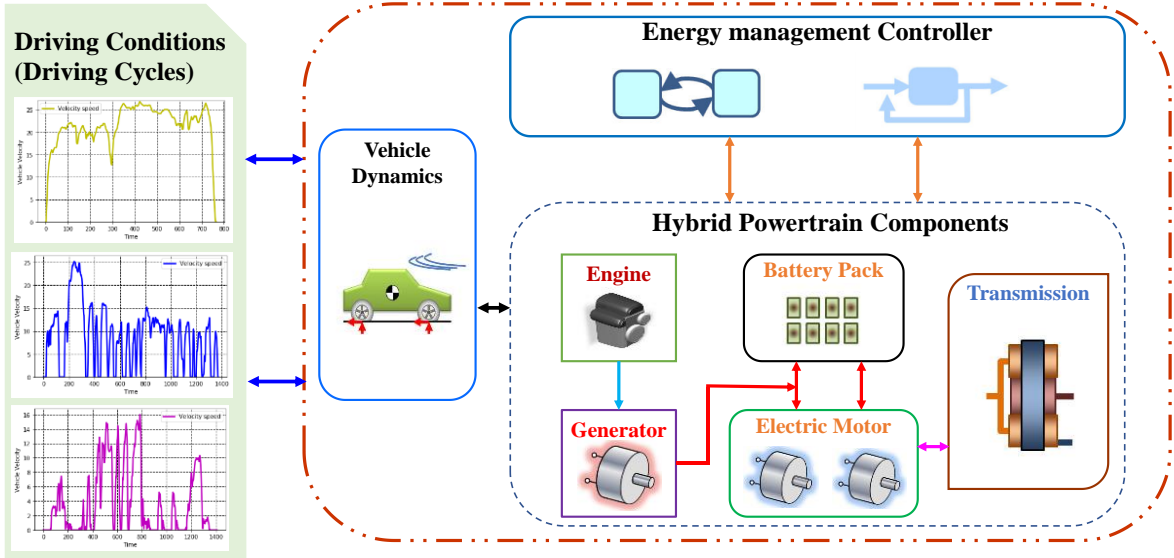

Fig. 1. Logical framework of driving cycle in energy management problem of HEV or PHEVs [20].

Motivated by the aforementioned literature gap, this review focuses on organizing a comprehensive survey on the driving cycle-driven EMSs for HEVs or PHEVs. First, the detailed list of multiple attempts regarding recent approaches for the energy management problem of HEV is introduced, including the review papers and regular articles on the advanced algorithms. The merits and demerits of these approaches are elaborated and compared in detail. In addition, the driving cycle-driven EMSs are divided into three categories according to the driving cycles' main features are completely known, partially known or totally unknown, as shown in Fig. 2. Based on this classification, the relevant research solutions, standard driving

| Categories / Features | Situation 1 | Situation 2 | Situation 3 |
|---|---|---|---|
| Travel Distance | ✓ | ✓ | ✗ |
| Duration Time | ✓ | ✗ | ✗ |
| Average Speed | ✓ | ✗ | ✗ |
| Average Acceleration | ✓ | ✗ | ✗ |
| Research Solution | Standard Driving Cycle | LT-DCG | ST-DCP |
| Potential Applications | Offline EMS | City bus, Sanitation truck, Delivery Vehicle | Taxi, Passenger cars |

✓ means the feature of driving cycle is known.
✗ indicates the feature of driving cycle is unknown.

Fig. 2. Three situations of driving cycle information in the energy management problem.

cycle, long-term driving cycle generation (LT-DCG), and short-term driving cycle prediction (ST-DCP) are emphasized and explained. Since the start and end points of the driving cycles are only known, LT-DCG could produce a whole driving cycle based on the collected historical data. However, ST-DCP focuses on forecasting the short-term horizon information according to current surrounding traffic conditions. The pros and cons of the correlated technologies are displayed and analyzed, which is beneficial to develop future research in the energy management field of hybrid vehicles. Furthermore, this survey article also concludes the effective databases of driving cycles for energy management research. The highway and urban driving environments are classified and the advantages and disadvantages of each database are labeled. As a result, the relevant researcher could choose an appropriate database for their own study expediently. This review hopefully accelerates the realization of real-time/online and adaptive EMS for hybrid vehicles in real-world environments.

The construction of the following paper is organized as follows. Three different power system structures of HEVs and the research status of algorithms in their energy management will be introduced in section 2. Section 3 will showcase the development of algorithms, information, and structures of HEVs' energy management system in the recent decade. The learning-based energy management method of HEVs was elaborated in section 4. Besides this effort, three types of EMSs, which depend on the driving cycle information is known or not, are introduced and studied in Section 5. In Section 6, the central open-source driving databases are evaluated based on the contained driving cycles information, and the future direction of the driving cycle's database is outlook. Finally, the conclusion of this review paper is presented in Section 7.

**2. Classified powertrains in HEVs**

In HEVs, there is more than one power resource jointly to supply the demand power. According to the different connection forms between components, the driver system of HEVs can be classified into three categories, which are series, parallel, and power-split, respectively. HEVs with different dynamic systems can operate in various driver modes and suitable for different driving conditions. Next, the structure features of each powertrain are introduced briefly, and then the research status of the corresponding framework was reviewed by several current pieces of literature.

*2.1 HEVs with series powertrain*

The structure of series HEVs is depicted as Fig. 3, and it is the simplest one of all frameworks. Series HEVs are electrically coupled driving methods that the ICE is not directly connected to the driveshaft, and the power demand of wheels is provided by a tractive motor. The power input into the motor is from battery and generator that is driven by ICE. Therefore, the size of the battery pack, motor, and generator is larger than the other two driver structures, and increased mass will bring additional fuel consumption. Compared to parallel HEVs, the engine is smaller in this structure. The intention of the engine is designed to extend the mileage of the vehicle and provide power in the situation of large required power as auxiliary energy. In conventional ICE vehicles, the ICE must work to meet the power demand from wheels. In many driving conditions, the ICE is running under the low efficiency. Especially in urban conditions, the car usually runs at a low speed, requires frequent starting and idling, the fuel consumption sharply increases in these situations. However, in series HEVs the ICE can keep running at the high-efficiency area to improve fuel economy. Thus, the series HEVs have good fuel-saving and emission-reducing performance under urban conditions. But in highway conditions, the engine needs to be turned on frequently to provide power, due to the loss of secondary energy transfer, series HEVs have higher fuel consumption than other types of vehicles.

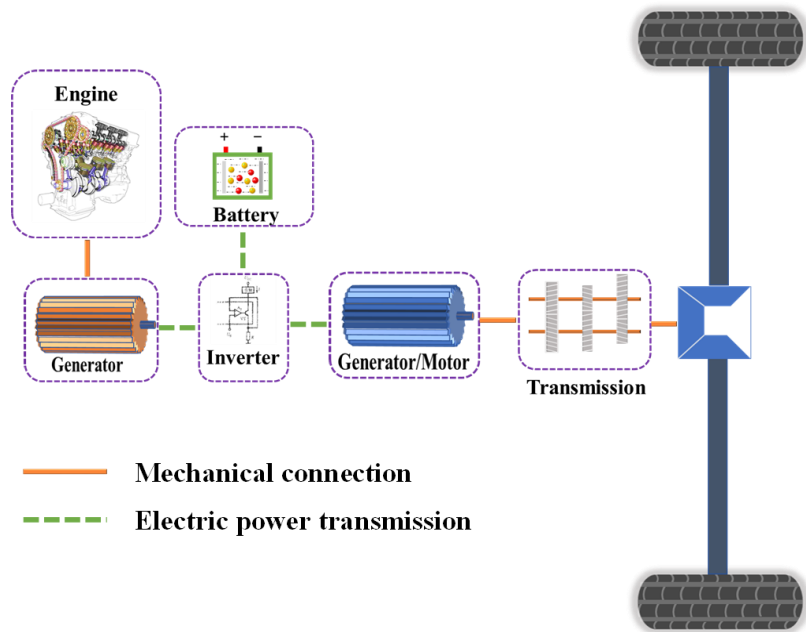

Fig. 3. The structure of series HEVs.

In the research of HEVs, EMS has always been a hot topic for researchers. In recent years, many pieces of research contributed to obtaining the optimal control strategy for series HEVs. For instance, Ref. [21] designed an online correction predictive EMS to acquire the optimal control policy for a series hybrid electric tracked vehicle (HETV), the multi-step predictor is applied to predict the future driving cycle and the dynamic programming (DP) is used to get the local optimal control action based on the future predicted driving cycle, the Q-learning algorithm adapted to reduce the impact of predict error online. However, the forecast error cannot be eliminated and the numerous amounts of calculation prevent its real-time online application. With the rise and application of artificial intelligence, the neural network (NN) is employed to address the issue of EMS for HEVs. The NN-based methods can process the problem with large amounts of data, so as to obtain more accurate results.

Deep reinforcement learning (DRL) is currently the most popular research method, which combines reinforcement learning (RL) and NN. The authors in [22] adopted a DQL algorithm in the power management problem of a series HETV, the simulation results show that the proposed method can reduce 5.96% fuel consumption than conventional RL algorithm. In Ref. [23], the group found that the Dyna-H algorithm can achieve more quickly convergence than Dyna, but the incompleteness training of the state-action pairs in Dyna-H will lead to bad adaptability. Meanwhile, the authors noticed that the DQL with AMSGrad optimizer can realize faster training speed than the conventional DQL with Adam optimizer. A dueling network was used in [24], using dueling architecture can distinguish whether the high reward is due to good driving conditions (e.g. states) or the execution of an action. Simultaneously, the prioritized experience replay was applied in the training progress to reduce the training time, the results indicate that the improved method can save about 3% fuel than original deep Q-network. Generally, the DRL algorithm can get an approximate global solution, but sometimes it will fall into overfitting. In [25], a modified double Q-learning was utilized to derive the optimal control for a series aircraft-towing tractor. Two heuristic action execution policies, the max-value-based policy, and the random policy, are employed to reduce the overestimation of the value function. By the experience of offline learning (software-in-the-loop) and online learning (hardware-in-the-loop), the results display the proposed way can save energy than general double Q-learning.

*2.2 HEVs with parallel powertrain*

A parallel HEVs power system is different from the series HEVs dynamic system, as shown in Fig. 4. There are two power transmission routes in the parallel HEVs power system. The first is that the engine is directly connected to the drive shaft through the transmission, and the other is that the battery pack is connected to the drive shaft through the traction motor. They can be independently or together to provide power for drive wheels. The engine is the main power source to power the vehicle, and the battery is used for braking energy recovery and auxiliary energy. Compare to series powertrain, the size of the battery and motor/generator is smaller, no generator is needed and only part of the engine power is converted by multi-power. Hence the overall efficiency is higher than the series structure and suitable for highway driving conditions. However, the parallel framework is mechanically coupled, and its control is more complicated than the series.

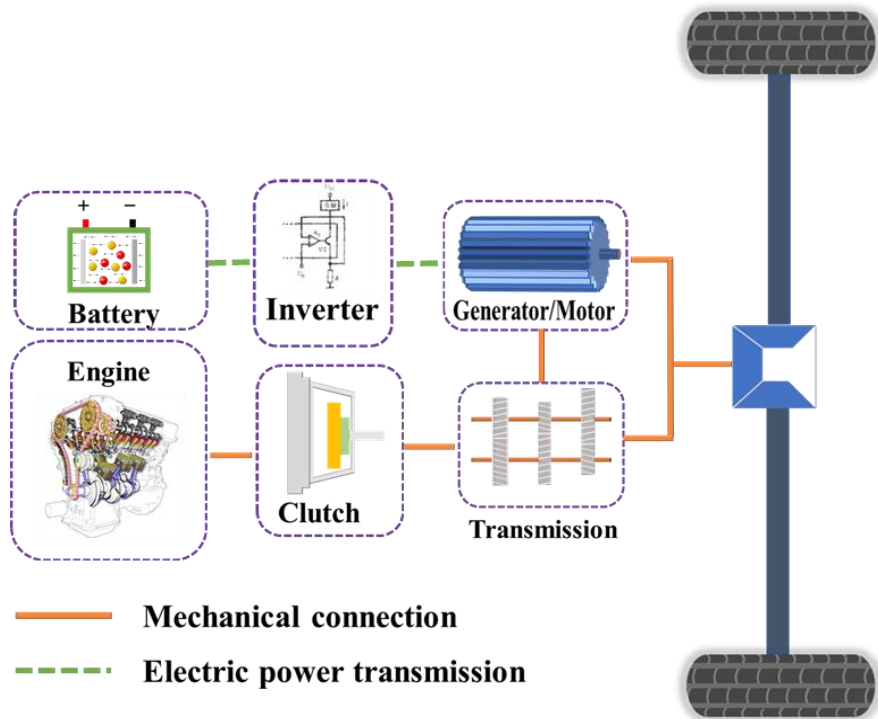

Fig. 4. The framework of parallel HEVs.

Parallel HEVs have been extensively studied due to their excellent fuel-saving performance, the RL algorithm has the ability of self-learning than the conventional optimal algorithm, so it is the ideal way to address the energy management for HEVs. The Q-learning algorithm was utilized to derive the optimal power split strategy for a parallel hybrid electric bus in [26], a linear radial basis function network was used to approximate the action-value function. It can achieve better performance of fuel economy and reduce emission than electric assistance strategy and model predictive control (MPC). The authors in [27] used the exponential weighting function to predict the demand power, and the auxiliary power consumption by automotive electronics such as light and air-condition is considered in the cost function, the TD (λ)-learning algorithm was employed to obtain the optimal strategy for a parallel HEV. The SOC is a very crucial variable and usually be selected as a state variable in the algorithm. But in [28], the SOC was excluded from state variables and take the policy that the battery is charged when the SOC reaches the lower boundary and discharged when it reaches the upper boundary. Compared with thermostatic rule-based strategy and ECMS, the designed method can reduce fuel consumption 8.89% and 0.88%, respectively.

Generally, traditional RL algorithms suffer from a large number of calculations and discrete errors of variables, which prevents their real-time application and gets accurate solutions. Fortunately, deep learning (DL) and transfer learning can be good at solving such problems. In [29], the author introduced three optimal algorithms, which are deterministic dynamic programming, stochastic dynamic programming (SDP) and Q-learning, respectively. The transfer learning method is proposed to address the issue of slow convergence for Q-learning. The method uses the optimal result of SDP to initialize the Q-value function of Q-learning and the consequence shows that this method can dramatically improve convergence speed than conventional Q-learning. In response to emissions issues, the method proximal policy optimization (PPO) was applied in [30] to reduce emission for a parallel HEV, which improves the efficiency of a selective catalytic reduction catalyst by controlling the temperature of exhaust gas. The bi-level algorithm structure was designed in [31], which contains the offline deep neural network phase and online DQL phase. In Ref. [32], the deep Q-network (DQN) was used for energy management of a parallel plug-in HEV, and the simulation results indicate the fuel consumption of DQN is higher around 6% than DP. Schroer et al. [33] adapted the deep deterministic policy gradient (DDPG) to train the RL agent to find an optimal energy management policy for a mild parallel HEV. The driver habits and traffic condition has been considered to reduce fuel consumption. Compared to DP, the applied algorithm can achieve near-optimal results.

*2.3 HEVs with power split powertrain*

Power-split powertrain is a drive system with speed coupling and torque coupling, as demonstrated in Fig. 5. Compared with the parallel framework, it has an additional motor working both as a generator and a tractive motor. At the same time, it uses a planetary gear mechanism to replace the traditional transmission. The overall mass is larger than the parallel type, so it needs to consume more energy. But its adaptability is better than the other two structures. It combines the merits of both series and parallel structure, it operates as a series of powertrain in low-speed conditions and parallel system under high-speed conditions. Because the series HEVs are more efficient under low-velocity conditions and parallel HEVs are more efficient in highway driving cycles.

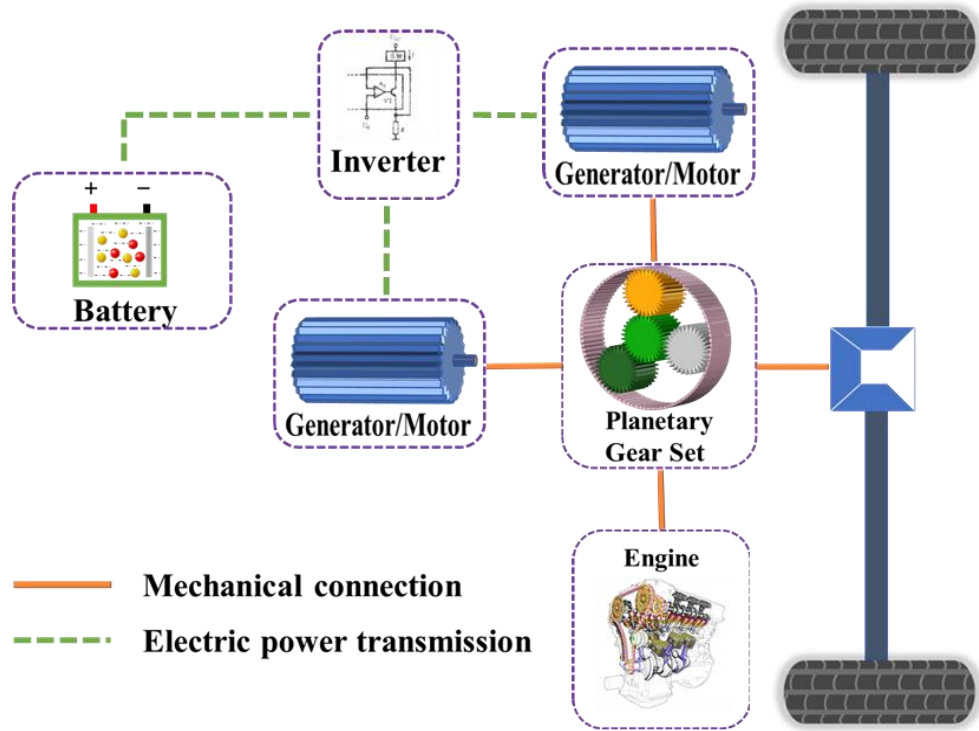

Fig. 5. The powertrain of power-split HEVs.

Due to its unique advantages, power-split HEVs have always been a research hotspot for researchers. In recent years, with the continuous improvement of RL algorithms, its application in the energy management of HEVs has achieved great results. The Q-learning is the most popular method for researchers in conventional RL algorithms. The author in [34] proposed a blended energy management control strategy for a power-split plug-in HEV, take the charge-depletion policy before the SOC reach a threshold, and then adopt the Q-learning algorithm. The adopted strategy can realize better fuel-saving performance than a conventional charge-sustaining strategy. In Ref. [35], the optimality of the Q-learning for energy management was studied in a power-split plug-in HEV, and the demand power transition probability matrix was calculated by maximum likelihood estimation. A rapid-DP method was developed in [36] by Yang et al. Firstly, a multi-criteria optimization method, Pareto optimality, was used to obtain the Pareto frontier operating points of the engine. In this way, the computation efficiency is greatly improved. Meanwhile, the joint optimization method that combined particle swarm optimization (PSO) and rapid-DP was designed to acquire the optimal control strategy and powertrain parameters for a power-split HEV (Prius).

The driving conditions have a significant impact in energy management for HEVs, the driving cycles were divided into short trips and long trips in [37]. The different algorithms QL-ST and QL-LT were employed in the corresponding driving cycle, and the optimal result of the QL-ST was used to initialize the QL-LT algorithm. The results indicate this way can reduce convergence time by about 70%. In [38], the researchers compared the performance of two algorithms, a greedy analytical control and Q-learning combined with neural dynamic programming. Based on short trips, the latter is better than the former. A self-learning structure Actor-Critic network was designed for a series-parallel plug-in hybrid electric bus to search for optimal EMS in [39]. Compared to conventional RL methods, this way can eliminate the discretization error and curse of dimensionality, its fuel-saving performance is better than discretization-based strategies. A DDPG algorithm that merged priority experience replayed and structured control nets were studied in [40], the environment is a plug-in hybrid electric bus, and the traffic information was embed in the training process to improve the optimal performance.

## 3. Overview of review literature for energy management strategies

Typically, the current literature on EMSs of hybrid vehicles contributes to two individual directions, the first one is comprehensive review papers in terms of different points of view and another one is regular articles aim to elaborate advanced algorithms and techniques in a particular energy management domain. Many researchers have proposed survey manuscripts according to various perspectives, such as the global optimization control methods, connected vehicles view, a combination of energy management and component sizing, specific MPC evolution for HEVs, etc. For example, Serrao et al. compared three optimal global approaches for energy management in 2011 [41], which are DP, Pontryagin's minimum principle (PMP), and equivalent consumption minimization strategy (ECMS). These methods-based results are served as the benchmark in the following years. The authors in [42] and [43] summarized and analyzed the EMSs in two independent topics, which are MPC-based ones and integrated ones considering power management and component size simultaneously. Owing to the diversity of the energy management methods, Ref. [44-47] conducted an extensive review of the existing algorithms for HEVs, wherein the three important powertrain architectures, as well as online and offline techniques, are discussed. Furthermore, Martinez et al. highlighted the EMSs in the context of connected vehicles and outlook on the future trends of PHEVs in the intelligent transportation system (ITS) [48]. Designing a through-the-road HEV with in-wheel motor is a popular concept in recent years, the authors in [49] concluded the pros and cons of this idea, and also compared its performance with the conventional HEVs. Aiming principally at a parallel hybrid electric vehicle, Enang et al. explained the realization process of different control methods on this configuration [50], including the workflow, equations, and parameters. A detailed overview of the review literature for energy management is depicted in Table 1. Although the driving cycles are highly significant for EMSs in HEV, few researchers have presented a comprehensive review of driving cycle-driven EMSs for HEVs.

The hotspot algorithms in the energy management field of HEVs or PHEVs are typically classified into two categories, online and offline EMSs. The offline EMSs are usually formulated based on the optimizing control theory, which indicates that they require the particular driving cycle information in advance. Based on the existing literature, they often carry heavy computation burden, and thus they are usually derived offline, however, they can be treated as a benchmark to validate the optimality of other methods. The relevant offline algorithms are DP, PMP, ECMS, genetic algorithm (GA), bee algorithm (BA), simulated annealing (SA), PSO, convex programming (CP), game theory (GT), etc. For example, PSO was first used in [51] to search the global power split controls. The consuming time and sub-optimal snare prevent it from a wide-range application. Different attempts of ECMS are executed in [52] and [53], and they are named as adaptive ECMS (A-ECMS) and telemetric ECMS (T-ECMS), in which the tuning rule of co-states is the critical point in this method. DP is one of the most common algorithms for optimal global controls [54-56], and its related results are often utilized to evaluate other novel techniques. Recently, Ref. [57] and [58] tried to extract the DP-correlated control criterions for real-time applications, however, the dependency of driving cycles cannot be overcome easily. By defining the appropriately initial population and tuning parameters, GA has the ability to carry out an optimal global search [59]. SA and BA have a faster convergence rate than GA [60], [61], however, they may trap in local optimum due to the enormous state space. Moreover, the authors in [62] and [63] applied CP to solve the energy management problem of fuel cell HEV and considered engine start and gearshift cost. The convex modeling of the powertrain is the pivotal point in this approach and it is not easily extended to the complicated powertrain. GT is suitable to manage the interaction between two agents, and thus Ref. [64] and [65] employed this method to handle the energy management and charging strategies for PHEVs. Up to this point, the offline algorithms could be leveraged in different hybrid powertrain and energy management problems, however, the necessity of the prior driving cycle information stops them from being applied in real-world environments (since this information is unknown in the real-world traffic).

Table 1. Content of current review papers in HEV's energy management.

| References | Powertrain Architecture | Content Description |
| --- | --- | --- |
| [41] | Series HEV | Describes and analyzes DP, PMP, and ECMS |
| [42] | HEVs and PHEVs | Elaborates MPC-based power management strategies and future study |
| [43] | All types of HEVs | Discusses factors that affect the performance of EMS and component sizing |
| [44-47] | All types of HEVs | Classify EMSs into online and offline types, and explain their pros and cons |
| [48] | PHEVs in connected environments | Highlight benefits of ITS, traffic information and cloud computing in EMSs |
| [49] | Through-the-road HEVs | Analyzes concept of HEVs with in-wheel motor and the related EMSs |
| [50] | Parallel HEV | Explains the realization process of popular algorithms in energy management field |

The online EMSs are further divided into two categories according to the time evolution, which are rule-based and instantaneous control ones. The rule-based algorithms are always dependent on the human experiences or the engineering knowledge, and they are represented as some certain criteria of some arguments, such as the engine torque, SOC in the battery, and the speed of the generator set. To enhance the performance of these rule-based EMSs in fuel economy and pollutant emissions, many researchers have proposed some advanced methods to generate rules, such as fuzzy-logic rule [66], power follower policy [67] and on/off strategy

Table 2. Comparative analysis of main algorithms in the energy management field.

| Algorithms | References | Categories | Description of Characteristics |
|---|---|---|---|
| DP | [54-56], [57, 58], [4], [16] | Offline EMS | Global optimality, dependency of driving cycles |
| PMP, ECMS | [52, 53], [4] | Offline/Online EMS | Key point is tuning co-states to accommodate driving conditions |
| GA | [59], [9] | Offline/Online EMS | Global optimality, defining initial population and parameters are core |
| BA | [61] | Offline EMS | Global optimality, better convergence rate and worse results than GA |
| SA | [60] | Offline EMS | Time-consuming, fluctuation of performance is large |
| PSO | [51] | Offline EMS | Suitable for multi-goals with random search strategy |
| CP | [62, 63] | Offline EMS | High requirement in convex modeling, less computation burden |
| GT | [64, 65] | Offline EMS | Time-consuming, high dependency on modeling construction |
| Fuzzy logic rules | [66] | Online EMS, rule-based | Online achievement, performance far from global optimum |
| Power follower | [67] | Online EMS, rule-based | Online achievement, requirement in special driving situations |
| On/off strategy | [68] | Online EMS, rule-based | Online achievement, worse than optimization-based results |
| SDP | [69-71] | Online EMS, instant.* | Data dependency, time-consuming, online EMS |
| SMC | [72] | Online EMS, instant. | Less applications, reference trajectories are necessary |
| NN | [73], [83] | Online EMS, instant. | Data dependency, high performance, tuning of parameters |
| RL | [74, 75], [76-78] | Online EMS, instant. | Real-time implementation, requirement in high-quality controller |
| Deep RL | [79-82] | Online EMS, instant. | Founded in artificial intelligence, popular methods, real-time EMS |

*Instant. indicates instantaneous power split controls.

[68]. Unfortunately, there is a lot of space in control performance to fill up when compared with the optimal global technologies. As alternatives, instantaneous control algorithms could achieve better performance while obtaining online control implementation. The related algorithms to instantaneous energy management policies are MPC, SDP, sliding mode control (SMC), NN, extremum seeking control (ESC), RL, etc. For example, many trials have conducted in MPC directions, such as stochastic MPC [69], nonlinear MPC [70], and linear varying-time MPC [71]. To effectively apply MPC in energy management problems, the high prediction and modeling accuracy are necessary. The authors in [72] considered the battery and super-capacitor in fully-active HEV's energy management, and a sliding-mode controller is built to control their currents to reference values. As the hybrid city bus has a regular route, an NN-based network is designed to train the length ratio and achieve online control [73], the related results can be served as a sub-optimal strategy.

## 4. Overview of learning-based energy management strategies

Learning-based EMSs founded in artificial intelligence are more and more popular in recent several years. These learning methods include supervised learning, unsupervised learning, reinforcement learning, deep learning, deep reinforcement learning, and so on. For example, the authors in [74] and [75] examine the optimality and adaptability of RL-based EMSs via comparing with the DP algorithm. The online RL-based power management policies integrated current and predictive driving cycle information are introduced in [76-78]. To fuse huge driving data to adapt to various driving situations, DL and RL are combined to derive online EMSs [79-82]. Specifically, Ref. [83] adopted a deep neural network (DNN) to train the action-value function in the RL framework and employed the Q-learning algorithm to compute the online controls. As a result, the obtained controls are free of the powertrain modeling and driving cycles. The authors in [84] constructed DRL-enabled power split controls based on stochastic driver models, and thus the results showed great potential to improve intelligence and adaptability. Furthermore, the energy efficiency of PHEV is able to be enhanced in ITS by sharing the real-time traffic conditions (driving cycles) with wireless communication, a global positioning system (GPS) or geographical information systems (GIS) [85-86]. A comprehensive overview of the different kinds of algorithms in the energy management field is displayed in Table 2. As the real-time applications of online EMSs are becoming more and more critical for HEVs, the access technologies of driving cycle information are especially significant for energy management, however, few references have summarized this topic meticulously.

*4.1 RL-based energy management strategies*

An optimal control strategy is very important to improve the performance (e.g., Fuel economy, drivability, emission reduction and so on) for HEVs, the problem of energy management in HEVs has always been a hotspot of research. In the early years, the rule-based methods and optimal-based methods are widely applied in the research of power split for HEVs. But the rule-base policy requires a wealth of engineering experience to formulate reasonable rules, and the optimization-based approach needs advanced knowledge of driving conditions. These constraints lead to poor performance in real-time applications. The RL algorithm is a self-learning method that has the potential to solve the previous problem. The main idea of the RL is to maximize the numerical return through the interaction between the agent and the environment, thereby obtaining the optimal control strategy. The framework of the RL algorithm as exhibited in Fig. 6.

Fuel cell hybrid electric vehicle (FCHEV) is considered as the cleanest transportation, which can minimize emissions. Usually, its power source includes a fuel cell and battery pack, many researchers are committed to reducing the consumption of hydrogen fuel in fuel cells through RL algorithms. In Ref. [87], the researchers employed the intelligent algorithm Q-learning to derive an optimal control strategy for a FCHEV, it aims to minimize the fuel consumption of fuel cell and reduce the SOC fluctuation of battery to improve the lifetime of the battery. The driving mode was divided into safe battery mode and economy mode in [88], and match to different reward functions. Compared with fuzzy logic control, the RL method has a better fuel economy. Combining the virtues of multiple algorithms is a good way, a hierarchical framework was developed to acquire the optimal control of power split for a FCHEV in [89]. The structure contains an adaptive fuzzy filter, an ECMS, and a Q-learning, the first two methods are applied to reduce the action space and the Q-learning was used to obtain the optimal strategy in real-time. The number of fuel cell stack start-stop was considered in [90], the RL and upper confidence tree search were utilized to optimize the control for a plug-in FCHEV. Compared to the rule-based strategy, the results demonstrated that hydrogen consumption reduce 6.14% and the start-stop times decrease significantly. In Ref. [91], the authors used a low-pass filter to reduce the start and off numbers of fuel-cell stack. Aim to prolong the service time of power sources and reduce hydrogen consumption. In [92], the SARSA algorithm was applied to address the issue of EMS for a FCHEV, the reward function was modeled as Gaussian distribution and the degree of hybridization was chosen as the action variable. A recursive algorithm was used to online update the TPM of demand power in [93], and the cosine similarity of two TPM decides whether to update the control policy. Meanwhile, the effects of the learning rate, discount rate, cosine similarity and forgetting factor on the performance of the algorithm are discussed separately. Finally, the optimal strategy acquired by Q-Learning for a plug-in FCHEV and the real-time performance was verified by the hardware-in-loop experiment.

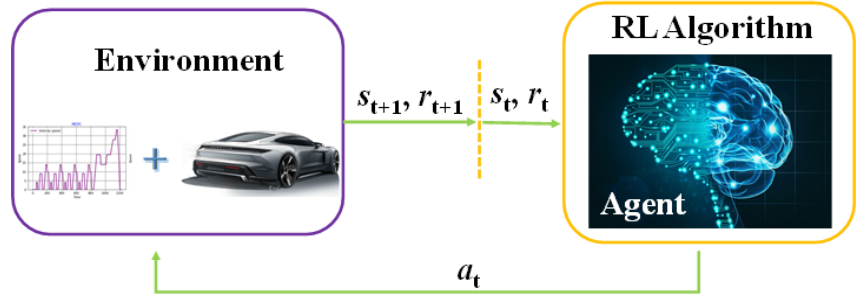

Fig. 6. The training process of RL.

Q-learning is an off-policy temporal-difference control algorithm, it is the most used RL method in the power management problem of HEVs. In [94], the maximum likelihood estimation and the nearest method were applied to derive the power demand TPM. Then, a recursive algorithm was employed to update the TPM online and the Q-learning used to optimize the control policy. The simulation result shows that the fuel consumption reduces 6% than stationary strategy. A fast Q-learning algorithm that used the Q-value of the previous time-step and the next time-step to update the Q-function was designed to obtain the optimal EMS for a HETV in [95]. At the same time, in order to adapt different driving cycles, the recursive algorithm and the KL divergence were used to update the control strategy online. The algorithm framework which combined the Q-learning and PMP was designed in [96], the optimal co-state obtained by the off-line training of PMP, then the optimal co-state as the action variable for Q-learning to get the optimal control strategy. This way can save 20.42% fuel consumption than charge-deleting and charge-sustaining control policy. In [97], an online EMS was obtained by combining the Q-learning algorithm and DP. The setting of hyperparameters has a great effect on the optimization performance of the algorithm, the researchers in [98] systematically analyzed the influence of hyperparameters on the fuel economy of HEVs, including the learning experience selection, the number of states, the discretization of states and actions, and the allocation between the exploration and exploitation.

In addition to Q-learning, many scholars have tried other RL algorithms in the energy management of HEVs. For instance, the author in [99] attempted to apply the policy iteration on the energy management for a super-mild HEV. the PMP algorithm was applied to obtain the analytical solution as the initial condition, and then the RL algorithm was employed to derive a real-time optimal control strategy for a hybrid excavator [100]. An integrated framework was proposed in [101], the first part of the structure is the off-line training use DP, then the optimal results derived from DP were embedded into the ECMS to optimize the power split strategy online. The performance of Q-learning with different numbers of state variable was studied in [102], the result indicates that the Q-learning with 4 state variables is better than others. The author in [103] adopted the Dyna-H method to realize the optimal control for a parallel HEV, compared with Q-learning, the model learning and planning process was inserted in the Dyna-H structure. Multiple algorithms were applied to obtain real-time EMSs for a PHEV in [104], the short-horizon driving pattern was predicted by a four order Markov chain speed predictor firstly. Then, the global optimal engine working point obtained by GA. the radial basis function neural network was applied to update the engine working point online to get the real-time EMS. The information of the V2V and V2I environment was employed as a part of state variables for the training of the PPO algorithm, and the local controller was utilized to improve the learning process by correcting the bad actions [105], Table 3 describes a detailed overview of the literature on the application of RL algorithms to the energy management of HEVs.

Table 3. Content of current literature uses the RL algorithms in HEVs' EMS.

| Algorithms | References | Content Description |
|---|---|---|
| Q-learning | [87-91], [93] | Derive the optimal EMS for FCHEV, aim to improve the FCHEVs' performance |
| SARSA | [92] | Compare the performance of Q-learning and SARSA in EMS for a FCHEV |
| Q-learning | [94,95] | Propose an improve Q-learning, embed the recursive algorithm to update the TMP online |
| Q-learning | [96,97], [100] | Combine the merits of Q-learning, PMP and DP |
| Q-learning | [98] | Analyzes the impact of algorithm hyperparameters on EMS |
| Policy iteration | [99] | Calculate the TPM of power demand, apply the EMS in real-time |
| DP | [101] | Employ the DP in off-line training and ECMS in the on-line application |
| Q-learning | [102] | Discuss the influence of the number of state variables in the Q-learning algorithm |
| Dyna-H | [103] | Analyzes the difference between the Dyna-H and Q-learning |
| Markov chain | [104] | Integrate the Markov chain, GA and radial basis function neural network to obtain a real-time EMS |
| PPO | [105] | Join the information of V2V and V2I in state variables, utilize the PPO to predict the future behavior |

Generally, for specific driving conditions, the RL method can converge well and obtain the global optimal solution. However, as the changes in the driving environment and conditions, its adaptability drops sharply. In addition, traditional RL algorithms cannot deal with continuous state space and continuous actions. It needs to discretize state and action variables, which inevitably brings discretization errors and cannot obtain accurate solutions. Moreover, as the discrete accuracy increases, the required calculation time and storage space increase exponentially. The previous constraints make it difficult to apply in real-time driving conditions. Fortunately, with the continuous research and application of DR, real-time application of EMSs has become possible.

*4.2 DRL-based energy management strategies*

The process of the DRL algorithm is displayed in Fig. 7. DRL algorithm combines DL and RL, which has greatly promoted the field of RL. With the help of NN, DL is good at handling massive amounts of data. For example, DQN is a currently mainstream and widely used DRL algorithm, which was proposed by researchers at Google DeepMind. It can reach the human level when playing any Atari game, and the recent AlphaGo intelligent robot defeated the world's human Go masters. These are enough to prove the strong adaptability of DRL algorithms to respond to real-time changing environments. The advent of the DRL algorithm solves the problem that traditional RL algorithms are prone to fall into the disaster of dimensionality due to discretization. At the same time, DRL is a model-free method with wide applicability and fast convergence speed. This has greatly promoted the online application of HEV EMSs. In recent years, many researchers have tried to apply DRL to the power management of HEVs, and have achieved good results.

In actual manufacturing and application, there are inevitably differences in the SOC of the battery, which is an important factor affecting the service life of the battery. The author in [106] applied the DRL approach to finding an optimal strategy to trade-off the SOC of all batteries, which can extend battery life and reduce maintenance. A method that combines RL and NN was proposed in [107] and used it to control the power allocation for an electric vehicle. The DQN method used to address the HEVs energy management in [108] and improves 3.51% fuel saving than rule-based strategy. In Ref. [109], the DQN algorithm was employed in the energy management of an extended-range electric vehicle. Compared with baseline vehicles, fuel consumption fell by 19.5%. In [110], the DQL was utilized to solve the power split issue of a hybrid electric bus, the simulation results indicated that the adopted method has good adaptability, optimality and faster convergence than Q-learning. These research results show that the DRL algorithms are better than the traditional RL algorithms in the application of HEVs EMS. This is because the continuous state variables can be used as input in DRL approaches, thereby reducing discrete errors and making the results more accurate. On the other hand, in the original RL algorithms, the Q-function is represented by the Q-value table, and it takes a lot of time to search for the optimal solution in practical applications. In the DRL algorithm, the Q-function is fitted by a DNN. As long as the optimal network parameters are found through offline training, the trained network is equivalent to a black box, one input can immediately get the optimal output. Thereby can greatly reduce the calculation time.

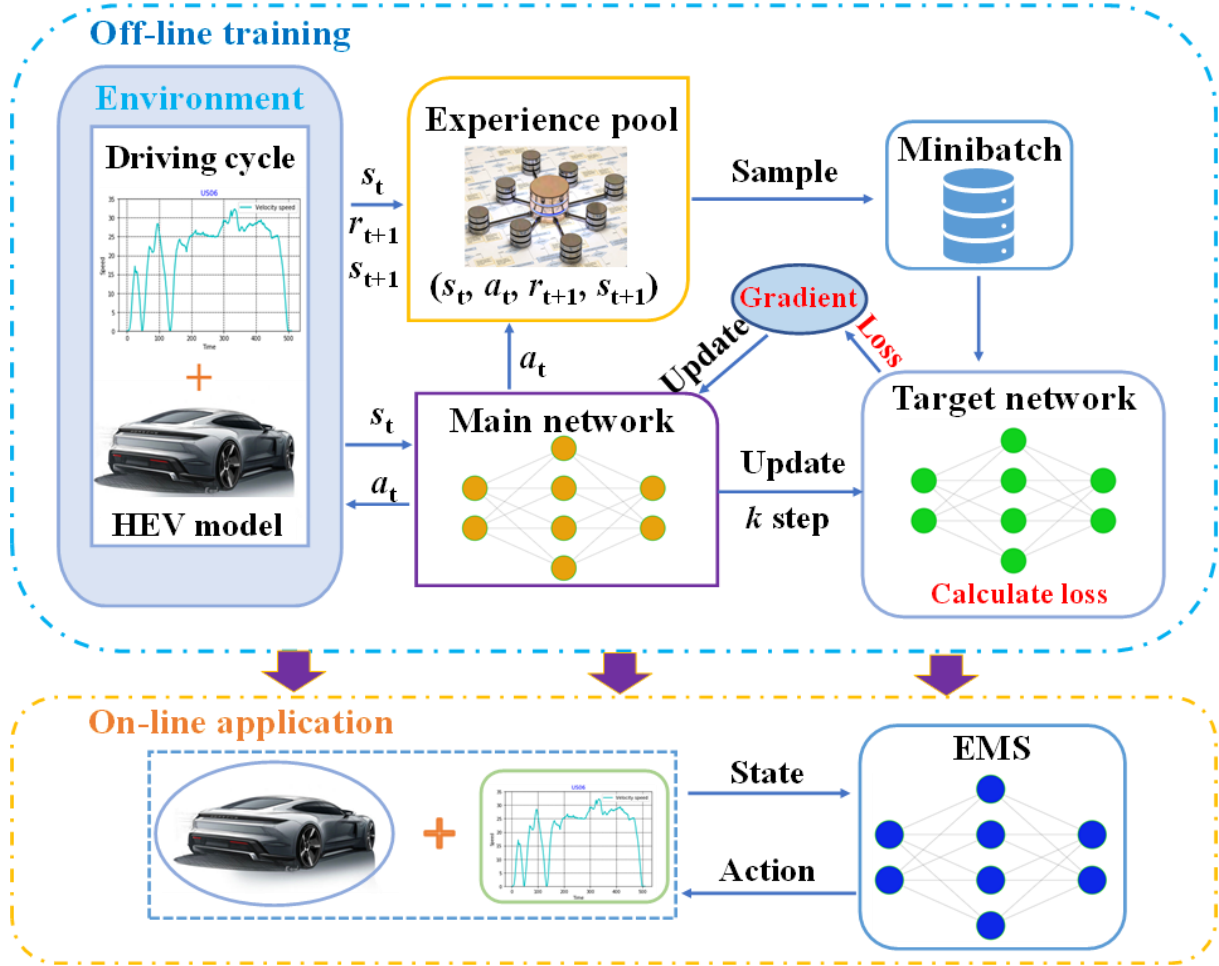

Fig. 7. The flow chart of DRL algorithm.

However, in the DQN algorithm, since the estimated Q-value and the target Q-value are calculated using the same NN, it is easy to overestimate the Q-value. Here are some tricks to solve previous problems. The first is to use the experience pool to store the training data, and then randomly extract a minibatch of data from it to train the NN every time, which can reduce the correlation between the data. The second is to use a double DQN to calculate the estimated Q-value and target Q-value separately, and the two networks use different parameters. A double deep Q-learning framework was designed in [111], which was adopted to derive an optimal EMS for a HETV. Compared with DQN, it can improve fuel economy and has better robustness and faster convergency. In [112], a dueling network structure was proposed in comparison to the DQN, the results suggested that it can reduce the time of convergence. The researchers in [113] found that the hyperparameters of hardware and algorithm can seriously affect the fuel consumption of HEVs. So, they employed Bayesian Optimization to optimize the hyperparameters such as the battery capacity, learning rate and electric machine dimension. Then the DRL used to acquire the optimal control strategy.

In the DQN algorithm, because the output of the NN is the probability value of each action or the corresponding Q-value, it cannot deal with the problem of continuous action space. But in the energy management of HEVs, the action variables are continuous values, such as engine output power, throttle opening, and so on. In response to this problem, many researchers have applied the DDPG algorithm to the energy management of HEVs and have made good progress. In [114], the DDPG was utilized in the power management of an extended-range electric vehicle and the fuel consumption is 21.8% lower than baseline vehicles. The author in [115] applied the DDPG algorithm to search the optimal EMS for a series HEV, it can achieve better performance than MPC. In order to speed up the computation, the critic network of DDPG adopted the dueling architecture that includes two streams of state value and action advantage [116]. Meanwhile, the real-time topographic information was applied in the training progress. In Ref. [117], three model-based methods which are Gaussian Process, Random Forest and Gradient Boosted Random Trees were utilized to optimize the hyperparameters of the DDPG algorithm. The results show that the Random Forest can obtain better hyperparameters to improve the performance of DDPG. To reduce the action space, the optimal brake specific fuel consumption curve of ICE and the power batter charge-discharge characteristics were embedded in the training of the DDPG algorithm in [118]. The safe framework was proposed in [119], the shield module was embedded between the environment and agent to protect the environment from inferior actions. Table 4 briefly summarizes the application of DRL algorithm in HEVs' energy management in recent years.

Table 4. Summary of the DRL algorithm literature in HEVs' EMS.

| References | Algorithms | Content Description |
|---|---|---|
| [106,107] | ANN | Find the optimal control strategy to allocate the power among the sources |
| [108-110] | DQN | Compare the DQN with traditional RL methods and rule-based strategy |
| [111] | Double DQL | Analyze the merits of double DQL than DQL |
| [112] | Dueling DQN | Design a dueling network structure to output the state value and action advantage severally |
| [113] | DRL | Use the Bayesian Optimization to optimize the hyperparameter of hardware and algorithm |
| [114] | Actor-Critic | Verify the advantage of Actor-Critic network by some comparative experience |
| [115-119] | DDPG | Compare DDPG with other methods, join some tricks to improve the optimal performance |

## 5. Driving cycle-driven EMSs for hybrid vehicles

From Fig. 2, the driving cycle information in the energy management field is always located in three situations, which are the main features that are completely known, partially known, and totally unknown. To handle each of these situations, different kinds of solutions are proposed, and they are standard driving cycles, LT-DCG, and ST-DCP. In this section, a comprehensive review of significant relevant algorithms on these three technologies is constructed. To the best of our knowledge, this is the first attempt to list the acquired approaches for driving cycle information in the energy management problem of HEVs or PHEVs.

### 5.1 Standard driving cycle for optimal global controls

As mentioned above, the offline EMSs require complete driving cycle information, including duration time, travel distance, average acceleration, and average speed. As a matter of fact, the United States, Europe, Japan, and China develop their own standard driving cycles for research and testing in the transportation field. In the energy management problem of hybrid vehicles, the standard driving cycles are usually utilized to evaluate the optimality of the EMSs and compare the control performance of different control technologies. For example, the authors in [82] applied two standard driving cycles, Urban Dynamometer Driving Schedule (UDDS) and New European Driving Cycle (NEDC) to estimate a heuristic planning RL-based energy management strategy and expound the importance of planning steps. Chen et al. aim to evaluate optimal global performance for DP, and thus they used many standard driving cycles to represent classic driving patterns, such as FTP75, J-1015, Taipei urban, and NYCC [56]. Furthermore, Onori et al. proposed an adaptive-PMP EMS in [120], and they integrated uphill and downhill into the standard driving cycles to see the variation of co-states. However, in real-world driving environments, the driving cycles are partially known or totally unknown, e.g., in the city bus, sanitation vehicles or taxis. Researchers or engineers should generate or predict the uncertain driving conditions information for the energy management controller. Based on this generated or predicted information, the controller is capable of formulating online EMSs to accommodate future driving situations. As a result, the hybrid vehicle could persistently improve fuel economy and reduce exhaust emission in different driving conditions.

*5.2 LT-DCG for partial information*

Two unavoidable reasons propel researchers to generate suitable driving cycles in energy management problem, wherein the first one is that the standard driving cycles are not appropriate for some hybrid powertrains, such as the collection truck and city bus. The second for driving cycle generation is that the main features of real-time driving cycles are often partially known (as the second situation in Fig. 2). Since the average speed and acceleration of one driving cycle are not determined, various types of driving cycles are able to be decided. At this stage, the most appropriate one should be generated and selected for energy management research according to current traffic conditions. For this purpose, many techniques have been presented to generate driving cycles, such as statistical distribution, learning vector quantization (LVQ), Markov chain (MC), evolutionary algorithm and mean tractive force (MTF). For example, the authors in [121-123] applied a stochastic and statistical methodology to generate a driving cycle based on long-range real-world driving data. Specifically, to create a representative driving cycle for passenger cars in Singapore, Ho et al. designed 12 road routes to collect the characteristics in the distance, road type, peak-lull proportion and duration [121]. Ref. [123] created more than 5000 driving cycles for a collection truck, and the most applicable one is chosen for energy management according to average power and time of one turnaround. Markov chain is usually utilized to mimic the driving cycle, which means the next vehicle speed depends on current speed and independent of the past history [124, 125]. In Ref. [126], 500 real-world driving cycles were processed as transition probability matrices (TPMs), which are further used with the Markov chain to generate stochastic driving cycles.

Besides the above-mentioned efforts, LVQ, MTF, and evolutionary algorithms are also leveraged to create driving cycles. For example, Perhinschi et al. applied a customized evolutionary algorithm to select micro-trips from many collecting trips based on some important parameters, such as average speed, stops per mile, and kinetic intensity [127]. The resultant software is easily realized in Matlab and could generate a reliable driving cycle in real-time for transportation research. The authors in [128] presented a bi-level control framework to obtain EMS, the upper-level realized driving cycle recognition and generation through LVQ, and the lower-level employed fuzzy controller to distribute the torques of engine and motor. The related control architecture is described in Fig. 8, wherein the inputs of the upper-level are the maximum speed, average speed, and low-speed factor to generate the driving cycle. This generated driving cycle is then the input of the lower-level to computed the optimal energy management policy based on fuzzy torque distribution. Furthermore, Nyberg et al. in [129] and [130] developed MTF components according to the Markov chain to generate driving cycles. MTF is defined as the vehicle's tractive energy at wheels in a special driving cycle, divided by the traveled distance. This method is evaluated as a sound engineering tool for testing and development. For HEV's energy management problem, LT-DCG can not only make the EMS match current driving situation better but also acquire more real driving cycles for different hybrid powertrains.

*5.3 ST-DCP for unknown situation*

In some driving tasks, the driving cycle information is absolutely unknown to the vehicular controller. The destination, duration, and speed trajectory are all random and uncertain (as described in the 3$^{rd}$ situation in Fig. 2). In this situation, assuming the future driving information can be acquired in advance, the onboard energy management controller could adjust the power spilt controls among multiple energy resources more reasonably. Therefore, the prediction driving cycle information is extremely essential for the formulation of EMS, especially the driving conditions are switched (e.g., from the highway to urban). With this goal in mind, many algorithms have been proposed to forecast driving cycles in HEV's energy management problem, such as *k*-nearest neighbor (*k*-NN), artificial neural network (ANN), GPS, Markov chain (MC), particle swarm optimization (PSO) algorithm, support vector machine (SVM), etc. For example, the authors in [131-133] used MC to simulate the driving cycles and then combined with particle filter or fuzzy logic rules to derive the predictive EMSs. The relevant experiment tests indicate the ST-DCP could effectively improve the control performance in different cost functions. In an intelligent transportation system (ITS), road information is attainable with the help of GPS [134], such as time, trip distance, acceleration, and velocity. According to the historical data, the future driving cycle information could easily be obtained based on the search of the existing database [135]. For large-scale driving data, SVM is regarded as an efficient tool to recognize and classify the predetermined features, and then forecast the to-be vehicle velocity and road slope [136].

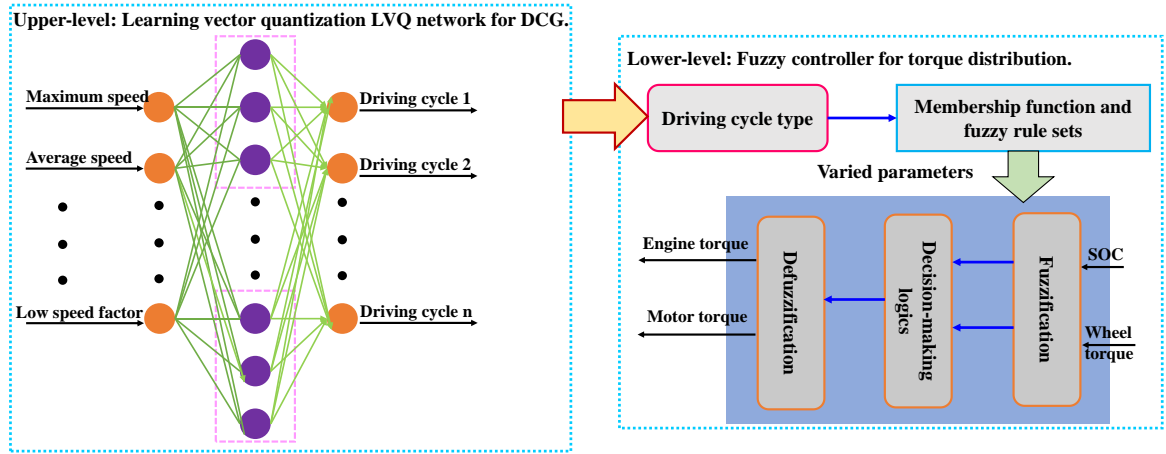

Fig. 8. A case study of LT-DCG in energy management field of HEVs [128].

Along with the deep learning method widely used in many scientific research areas, ANN is found as a promising solution to achieve precise prediction for time sequence (a driving cycle is able to be treated as a time sequence). For the energy management problem of HEV or PHEVs, the authors of Ref. [137-140] tried to apply different types of neural network (NN) to realize ST-DCP. For example, Feng et al. [137] and Xiang et al. [139] selected radial basis function NN to get the future power demand and vehicle speed, respectively. These novel predictive EMSs reflect better performance via comparing with ECMS and PID controls. In Ref. [138], the authors employed two ANN to address vehicle speed measurement and some relevant route information, respectively. The diagram of this driving cycle prediction block is sketched in Fig. 9, wherein pre-processing, prediction, and post-processing are included. Based on this predicted information, the potential of related HEV can be further improved. In addition, the author in [141] attempted a dynamic-neighborhood PSO algorithm to increase the accuracy of ST-DCP, as a result, the energy usage can be reduced up to 10%. Furthermore, Ref. [142] and [143] employed another two algorithms for ST-DCP, which are the car-following model and MPC. Based on the particular models, the future speed information can be easily obtained to promote the performance of many control objectives. By predicting the future driving cycles, the energy management controller is able to not only save fuel consumption but also enhance the safety and stability of the powertrain. For better observation and understanding, Table 5 exhibits the main approaches for LT-DCG and ST-DCP in HEVs' energy management field, which aims to emphasize the significance of random and future driving cycle information.

## 6. Overview of open-source driving cycle database

In this section, an extensive presentation of open-source driving cycle databases is conducted. These driving cycle databases are located in three different aspects, which are standard, highway, and urban driving cycles. In the energy management problem of HEVs, the standard driving cycles are usually adopted to evaluate the correctness and effectiveness of a novel EMS because all information on these cycles are known to researchers. The highway and urban driving cycles always contain random and uncertain features, and thus they are suitable to evaluate the online EMS in real-world environments. This section introduces the databases for these three kinds of driving cycles, which aim to provide the research basis for the study of optimal, online, and real-time EMSs in different driving situations.

*6.1 Standard driving cycles*

Standard driving cycles are often utilized to assess fuel consumption and pollutant emissions in a normalized way, especially for commercial vehicles. These cycles are performed on an engine dynamometer, and the performance is evaluated by a set of engine torque and speed points. Standard driving cycles in energy management problems are often modal cycles, which means they contain straight acceleration and constant speed periods [144]. The typical standard cycles are the New European Driving Cycle (NEDC), Federal Test Procedure (FTP)-75 cycle, Highway Fuel Economy Test (HWFET) cycle, Japan cycle'08 (JC08), etc [145].

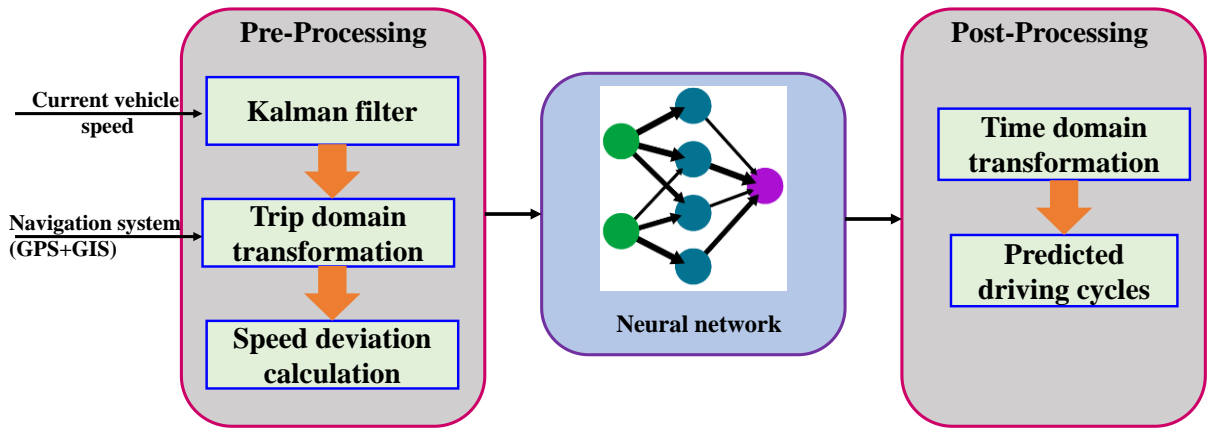

Fig. 9. A case study of ST-DCP in the energy management field [138].

Table 5. The current literature related to LT-DCG and ST-DCP algorithms in energy management problems.

| Algorithms | Literature | Realization Form | Characteristic Representation |
|---|---|---|---|
| Statistical distribution | [120-123] | LT-DCG | Necessary multiple features, data dependency |
| LVQ | [128] | LT-DCG | Driving style recognition and generation, bi-level controller |
| MC | [124-126] | LT-DCG | Computation of transition probability matrix, data collection |
| Evolutionary algorithm | [127] | LT-DCG | Main Parameters decision, real-time implementation |
| MTF | [129, 130] | LT-DCG | MTF components calculation and online application |
| MPC | [143] | ST-DCP | Predicted accuracy depends on estimated modeling |
| ANN | [137-140] | ST-DCP | Various types, time-consuming for training |
| Onboard GPS | [134, 135] | ST-DCP | Dependency of historical data, low robustness |
| MC | [131-133] | ST-DCP | Combined structure, particle filter or fuzzy logic |
| PSO | [141] | ST-DCP | Improve prediction accuracy, offline achievement |
| SVM | [136] | ST-DCP | Necessary predetermined features, speed and slope |

Fig. 10 depicts an illustration of standard driving cycles in Europe, China, the United States, and Japan. For example, NEDC includes two different parts, wherein the first part is four segments of the urban driving cycle (UDC) and the second part is an extra-urban driving cycle (EUDC). The critical characteristics of NEDC are, the traveled distance is 11023 m, the duration time is 1180 s, and the average speed is 33.6 km/h [146]. Besides the urban driving condition, standard driving cycles also refer to motorway conditions. The Artemis driving cycles are produced by the Artemis project in Europe, and these cycles include motorway 130 km/h and motorway 150 km/h [147]. The main features of the motorway 150 km/h cycle are, the distance is 29545 m, the time is 1068 s, and the average speed is 99.6 km/h.

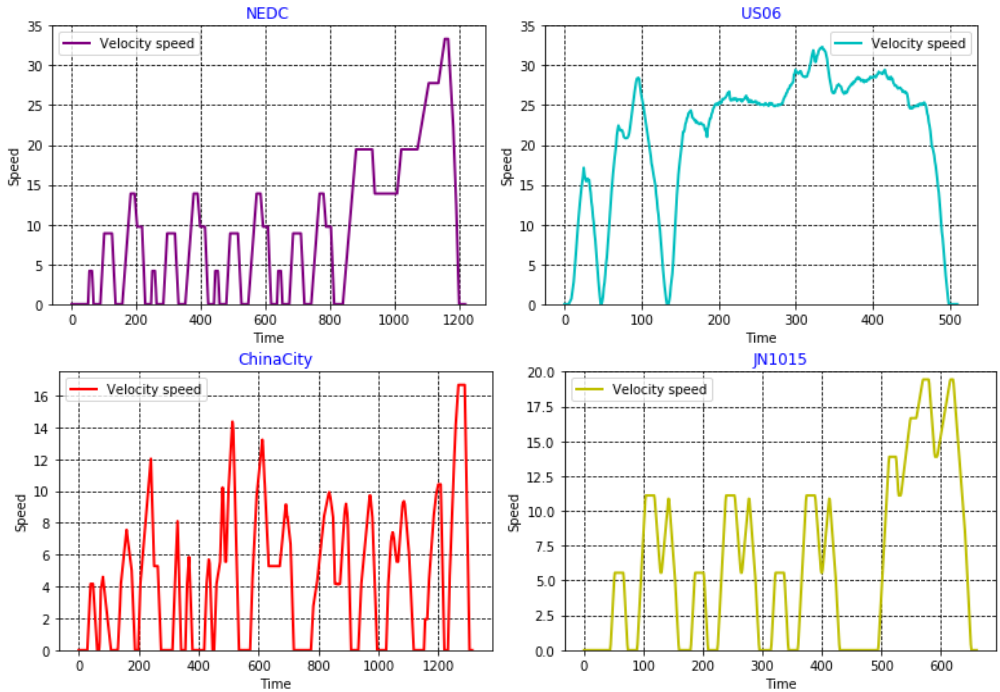

Fig. 10. Example of standard driving cycles in different areas.

To develop an EMS for the HEV and PHEV, it is important to certify this proposed policy with a standard driving cycle. Three factors determine the significance of the standard driving cycles in the energy management field. First, these cycles are generated professionally without noise, and thus it can guarantee the safety of the powertrain. Second, the standard cycle is unified in speed, acceleration, time, and distance, it is convenient to compare the control performance of multiple EMSs. Finally, these cycles could represent criterions in different areas all over the world, and they can help automotive manufacturers produce appropriate HEVs for various regions.

*6.2 Real-world urban driving cycles*

Different from the standard driving cycle, the real-world cycles are collected by on-road vehicles with installed sensors. The most common devices for data collection are GPS, cameras, Lidar, and radar. GPS is easily able to obtain the information of position, speed, acceleration, and distance for the vehicles. The other three sensors can be employed to acquire the knowledge of surrounding driving environments, such as other surrounding vehicles, pedestrians, lanes, and traffic lights. Hence, the data collected by on-road vehicles established GPS are enough for energy management research.

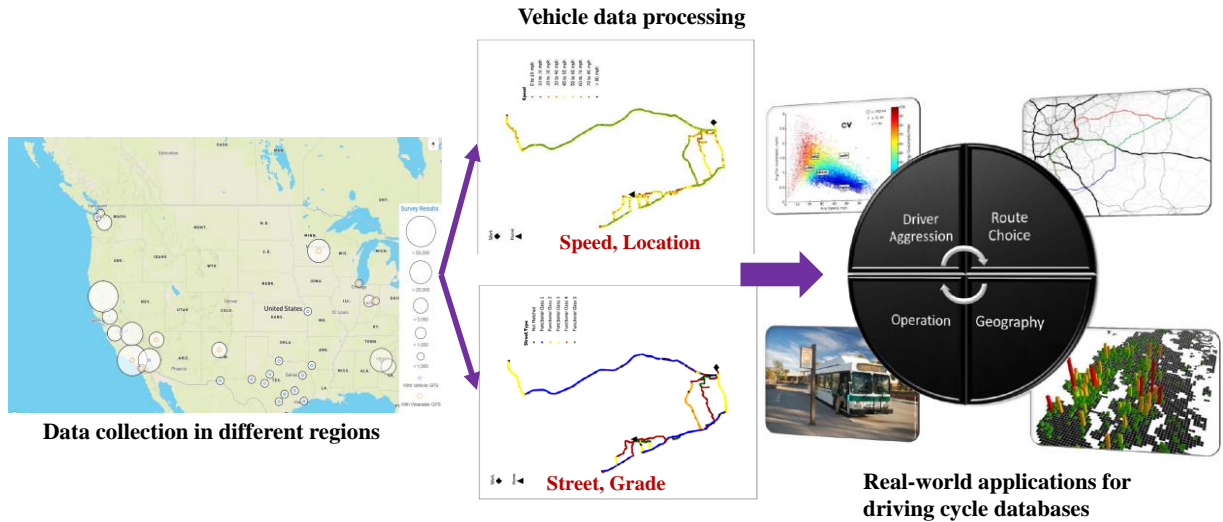

Fig. 11. Data coverage of the TSDC database in states and regions of the U.S [149].

For urban driving cycles, the Transportation Secure Data Center (TSDC) in the United States provide free, web-based access to detailed transportation data from a variety of travel surveys [148, 149]. The GPS data coverage in different states and regions provided by TSDC and its normal usage is shown in Fig. 11. These applications include geography, route choice, bus operation, and analysis of driver behaviors. Totally, TSDC hosted 35 surveys, and the traveled miles are more than 11 million, and the publications citing TSDC data are nearly 150. This data is appropriate for planners, researchers, and manufacturers and can be utilized in many applications, such as congestion mitigation, energy, and power analysis, alternative fuel station planning, and transit planning. This data supports many software, e.g., Spyder, Python, ArcGIS, and MS Office.

Specifically, the U.S. Department of National Renewable Energy Laboratory (NREL) provides second-by-second driving cycle data [150]. These cycles involve many travel surveys, such as transportation, household travel, and regional travel survey. They are mainly urban cycles, which include the information of speed and acceleration for the vehicles. Innovative algorithms are developed to provide optional routes for various types of vehicles to improve fuel efficiency. Moreover, these driving cycles are very beneficial for energy management research in hybrid vehicles.

*6.3 Real-world highway driving cycles*

One of the most famous and popular highways driving cycle databases is the Next Generation

Simulation (NGSIM) program [151]. This program was launched to collect vehicle trajectory data on four motorways in the U.S., which are eastbound I-80 in Emeryville, Peachtree Street in Atlanta, Lankershim Boulevard in Los Angeles and southbound US 101. The related data consist of vehicle speed, acceleration, vehicle type, lane number, position, vehicle length and width, and so on. Typically, there are 4 to 6 lanes in these highway databases. Different from the above-mentioned urban driving cycle databases, this one was collected via digital video cameras and customized software (named NGVIDEO) was developed to transcribe the vehicle trajectory data from the video [152]. These real-world and high-quality driving cycle datasets are utilized and researched in many fields, such as interchange configuration, traffic lights coordination, and intelligent transportation system.

Especially, the I-80 dataset [153] was occurred in the San Francisco Bay area in Emeryville, CA, on April 13, 2005. The architecture of the research area is about 500 meters and 6 highway lanes, and the sampling frequency is 10 Hz. Three 15-minutes segments are contained in this dataset, which are 4:00 p.m. to 4:15 p.m., 5:00 p.m. to 5:15 p.m., and 5:15 p.m. to 5:30 p.m. These periods include uncongested, congested driving conditions, and the mixture of them. For research purposes, the I-80 dataset is usually exploited to verify the algorithms of choosing lane on a highway, merging from on-ramp, and understanding driver behaviors [154]. As an example, the preview of the I-80 dataset tables is described in Fig. 12, in which the physical meaning of each column is explained in a given document.

The NGSIM database is also useful for energy management research in HEVs. For example, when the driving situation transforms from uncongested to congest condition, how to adjust the EMS quickly and efficiently to improve the fuel economy for these vehicles. Additionally, the aggressive driving style can be interpreted as that the drivers make lane-change and acceleration decisions frequently to save time, how to design suitable EMSs for these drivers or common drivers is an interesting and severe task. Moreover, assuming the vehicles on the highway are located in connected environments, how to tune the EMS for different vehicles if the driving information of the surrounding vehicles is known through connected communication technologies. These research directions can be realized and evaluated based on this NGSIM

Fig. 12. Overview of multiple parameters and variables in the NGSIM database.

Table 6. Optional driving cycles database for HEV's energy management research.

| Name | Resources | Form | Feature Description |
|---|---|---|---|
| NEDC, JC08, HWFET, etc | [144, 145] | Standard DC* | Modal cycles, contain straight acceleration and constant speed periods |
| TSDC and NREL | [148, 149] | Urban DC | Fresh driving data published in 2019, moderate quantity, multiple types |
| NGSIM | [150-152] | Motorway DC | Enormous quantity, published in 2006, four places, different driving condition |
| HighD dataset | [155] | Motorway DC | Published in 2018, moderate quantity, mainly high speed, German highway |

*DC: driving cycles

database. Finally, Table 6 lists the discussed driving cycle databases, which are appropriate for energy management research and may be beneficial for further study for online EMSs in this area.

## 7. Conclusion

Since driving conditions (especially driving cycles) are extremely significant in HEV's energy management problem, this paper aims to construct a comprehensive review of driving cycles-driven EMSs until now. The research background and current research progress of energy management in hybrid vehicles are introduced. Almost all the algorithms for offline and online EMSs are mentioned and analyzed. As the core of this review, the approaches for standard driving cycles, LT-DCG and ST-DCP are deeply discussed and compared to address the situations wherein the information of driving cycles is not known completely. In addition, different kinds of driving cycle databases appropriate for energy management study are introduced in detail, which hopefully help the researchers develop further studies in real-time and online EMSs for HEV and PHEVs.

**References**

[1] T. Liu, B. Wang, and C. Yang, "Online Markov Chain-based energy management for a hybrid tracked vehicle with speedy Q-learning," *Energy*, vol. 160, pp. 544-555, 2018.

[2] S. Ramachandran, and U. Stimming, "Well to wheel analysis of low carbon alternatives for road traffic," *Energy Environ Sci*, vol. 8, no. 11, pp. 3313–24, 2015.

[3] Z. Song, X. Zhang, J. Li, H. Hofmann, M. Ouyang, and J. Du, "Component sizing optimization of plug-in hybrid electric vehicles with the hybrid energy storage system," *Energy*, vol. 144, pp. 393-403, 2018.

[4] Y. Zou, T. Liu, F. Sun, and H. Peng, "Comparative study of dynamic programming and Pontryagin's minimum principle on energy management for a parallel hybrid electric vehicle," *Energies*, vol. 6, no. 4, pp. 2305-2318, 2013.

[5] F. Momen, K. Rahman, and Y. Son, "Electrical propulsion system design of Chevrolet Bolt battery electric vehicle," *IEEE Trans. Ind. Applicat.*, vol. 55, no. 1, pp. 376-384, 2019.

[6] M. Pourabdollah, B. Egardt, N. Murgovski, and A. Grauers, "Convex optimization methods for powertrain sizing of electrified vehicles by using different levels of modeling details," *IEEE Trans. Veh. Technol.*, vol. 67, no. 3, pp. 1881-1893, 2018.

[7] X. Wu, X. Hu, X. Yin, L. Li, Z. Zeng, and V. Pickert, V. "Convex programming energy management and components sizing of a plug-in fuel cell urban logistics vehicle," *Journal of Power Sources*, vol. 423, pp. 358-366, 2019.

[8] S. Xie, X. Hu, T. Liu, S. Qi, K. Lang, and H. Li, "Predictive vehicle-following power management for plug-in hybrid electric vehicles," Energy, vol. 166, pp. 701-714, 2019.

[9] T. Liu, H. Yu, and X. Hu, "Robust energy management strategy for a range extender electric vehicle via genetic algorithm," in *Proc. 2018 IEEE Vehicle Power and Propulsion Conference* (VPPC), pp. 1-6, August, 2018.

[10] M. Kim, and H. Peng, "Power management and design optimization of fuel cell/battery hybrid vehicles," *Journal of power sources*, vol. 165, no. 2, pp. 819-832, 2007.

[11] R. Johri, and Z. Filipi, "Optimal energy management of a series hybrid vehicle with combined fuel economy and low-emission objectives," in *Proc. Institution of Mechanical Engineers, Part D: Journal of Automobile Engineering*, vol. 228, no. 12, pp. 1424-1439, 2014.

[12] A. Castaings, W. Lhomme, R. Trigui, and A. Bouscayrol, "Comparison of energy management strategies of a battery/supercapacitors system for electric vehicle under real-time constraints," *Applied Energy*, vol. 163, pp. 190-200, 2016.

[13] S. Althaher, P. Mancarella, and J. Mutale, "Automated demand response from home energy management system under dynamic pricing and power and comfort constraints," *IEEE Trans. Smart Grid*, vol. 6, no. 4, pp. 1874-1883, 2015.

[14] L. Zhang, X. Hu, Z. Wang, F. Sun, J. Deng, and D. Dorrell, "Multiobjective optimal sizing of hybrid energy storage system for electric vehicles," *IEEE Trans. Veh. Technol.*, vol. 67, no. 2, pp. 1027-1035, 2018.

[15] R. Langari, and J. Won, "Intelligent energy management agent for a parallel hybrid vehicle-part I: system

architecture and design of the driving situation identification process," *IEEE Trans. Veh. Technol.*, vol. 54, no. 3, pp. 925-934, 2005.

[16] S. Zhang, and R. Xiong, "Adaptive energy management of a plug-in hybrid electric vehicle based on driving pattern recognition and dynamic programming," *Applied Energy*, vol. 155, pp. 68-78, 2015.

[17] S. Xie, X. Hu, S. Qi, X. Tang, K. Lang, Z. Xin, and J. Brighton, "Model predictive energy management for plug-in hybrid electric vehicles considering optimal battery depth of discharge," *Energy*, vol. 173, pp. 667-678, 2019.

[18] L. Li, C. Yang, Y. Zhang, L. Zhang, and J. Song, "Correctional DP-based energy management strategy of plug-in hybrid electric bus for city-bus route," *IEEE Trans. Veh. Technol.*, vol. 64, no. 7, pp. 2792-2803, 2015.

[19] J. Shen, and A. Khaligh, "A supervisory energy management control strategy in a battery/ ultracapacitor hybrid energy storage system," *IEEE Trans. on Trans. Elect.*, vol. 1, no. 3, pp. 223-231, 2015.

[20] S. Miller, "Hybrid-Electric Vehicle Model in Simulink (https://www.github.com/mathworks/Simscape-HEV-Series-Parallel)," *GitHub*. Retrieved June 12, 2020.

[21] J. Wu, Y. Zou, X. Zhang, T. Liu, Z. Kong and D. He, "An online correction predictive EMS for a hybrid electric tracked vehicle based on dynamic programming and reinforcement learning," *IEEE Access*, vol. 7, pp. 98252-98266, 2019.

[22] D. He, Y. Zou, J. Wu, X. Zhang, Z. Zhang and R. Wang, "Deep Q-Learning based energy management strategy for a series hybrid electric tracked vehicle and its adaptability validation," 2019 *IEEE Transportation Electrification Conference and Expo* (ITEC), pp. 1-6, 2019.

[23] G. Du, Y. Zou, X. Zhang, T. Liu, J. Wu and D. He, "Deep reinforcement learning based energy management for a hybrid electric vehicle," *Energy*, vol. 201, 2020.

[24] Y. Li, H. He, J. Peng and J. Wu, "Energy management strategy for a series hybrid electric vehicle using improved deep Q-network learning algorithm with prioritized replay," *DEStech Transactions on Environment Energy and Earth Science*, DOI: 10.12783/dteees/iceee2018/27794.

[25] B. Shuai, Q. Zhou, J. Li, Y. He, Z. Li, H. Williams, H. Xu and S. Shuai, "Heuristic action execution for energy efficient charge-sustaining control of connected hybrid vehicles with model-free double Q-learning," *Applied Energy*, vol. 267, 2020.

[26] Y. Fang, C. Song, B. Xia and Q. Song, "An energy management strategy for hybrid electric bus based on reinforcement learning," *The 27th Chinese Control and Decision Conference* (2015 CCDC), pp. 4973-4977, 2015.

[27] Y. Wang, X. Lin, M. Pedram and N. Chang, "Joint automatic control of the powertrain and auxiliary systems to enhance the electromobility in hybrid electric vehicles," *Proceedings of the 52nd Annual Design Automation Conference on - DAC '15*, pp. 1-6, 2015.

[28] B. Xu, F. Malmir, D. Rathod and Z. Filipi, "Real-time reinforcement learning optimized energy management for a 48V mild hybrid electric vehicle," *SAE Technical Paper* 2019-01-1208, 2019.

[29] H. Lee, C. Song, N. Kim and S.W. Cha, "Comparative analysis of energy management strategies for HEV: dynamic programming and reinforcement learning," *IEEE Access*, vol. 8, pp. 67112-67123, 2020.

[30] J. Hofstetter, H. Bauer, W. Li and G. Wachtmeister, "Energy and emission management of hybrid electric vehicles using reinforcement learning," *IFAC-PapersOnLine*, vol. 52, no. 29, pp. 19-24, 2019.

[31] P. Zhao, Y. Wang, N. Chang, Q. Zhu and X. Lin, "A deep reinforcement learning framework for optimizing fuel economy of hybrid electric vehicles," 2018 *23rd Asia and South Pacific Design Automation Conference* (ASP-DAC), pp. 196-202, 2018.

[32] C. Song, H. Lee, K. Kim and S.W. Cha, "A power management strategy for parallel PHEV using deep Q-Networks," 2018 *IEEE Vehicle Power and Propulsion Conference* (VPPC), *IEEE*, pp. 1-5, 2018.

[33] C. Schroer, R. Ließner and B. Bäker, "Adaptive operating strategies for the energy management of hybrid electric vehicles with deep reinforcement learning," Conference: *Electric & Electronic Systems in Hybrid and Electrical Vehicles and Electrical Energy Management* (EEHE 2018), 2018.

[34] X. Qi, G. Wu, K. Boriboonsomsin and M.J. Barth, "A novel blended real-time energy management strategy for plug-in hybrid electric vehicle commute trips," 2015 *IEEE 18th International Conference on Intelligent Transportation Systems*, pp. 1002-1007, 2015.

[35] Z. Chen, H. Hu, Y. Wu, R. Xiao, J. Shen and Y. Liu, "Energy management for a power-split plug-in hybrid electric vehicle based on reinforcement learning," *Applied Sciences*, vol. 8, 2018.

[36] Y. Yang, H. Pei, X. Hu, Y. Liu, C. Hou and D. Cao, "Fuel economy optimization of power split hybrid vehicles: A rapid dynamic programming approach, Energy," vol. 166, pp. 929-938, 2019.

[37] C. Liu and Y.L. Murphey, "Optimal power management based on Q-Learning and neuro-dynamic programming for plug-in hybrid electric vehicles," *IEEE Trans Neural Netw Learn Syst*, DOI: 10.1109/TNNLS.2019.2927531(2019).

[38] C. Liu and Y.L. Murphey, "Analytical greedy control and Q-learning for optimal power management of plug-in hybrid electric vehicles," 2017 *IEEE Symposium Series on Computational Intelligence* (SSCI), *IEEE*, pp. 1-8, 2017.

[39] H. Tan, H. Zhang, J. Peng, Z. Jiang and Y. Wu, "Energy management of hybrid electric bus based on deep reinforcement learning in continuous state and action space," *Energy Conversion and Management*, vol. 195, pp. 548-560, 2019.

[40] Y. Wu, H. Tan, J. Peng, H. Zhang and H. He, "Deep reinforcement learning of energy management with continuous control strategy and traffic information for a series-parallel plug-in hybrid electric bus," *Applied Energy*, vol. 247, pp. 454-466, 2019.

[41] L. Serrao, S. Onori, and G. Rizzoni, "A comparative analysis of energy management strategies for hybrid electric vehicles," *Journal of Dynamic Systems, Measurement, and Control*, vol. 133, no. 3, pp. 031012, 2011.

[42] Y. Huang, H. Wang, A. Khajepour, H. He, and J. Ji, "Model predictive control power management strategies for HEVs: A review," *Journal of Power Sources*, vol. 341, pp. 91-106, 2017.

[43] Y. Huang, H. Wang, A. Khajepour, B. Li, J. Ji, K. Zhao, and C. Hu, "A review of power management strategies and component sizing methods for hybrid vehicles," *Renew Sustain Energy Rev*, vol. 96, pp. 132-144, 2018.

[44] A. Panday, and H. Bansal, "A review of optimal energy management strategies for hybrid electric vehicle," International Journal of Vehicular Technology, vol. 2014, 2014.

[45] F. Salmasi, "Control strategies for hybrid electric vehicles: Evolution, classification, comparison, and future trends," *IEEE Trans. Veh. Technol.*, vol. 56, no. 5, pp. 2393-2404, 2007.

[46] A. Malikopoulos, "Supervisory power management control algorithms for hybrid electric vehicles: A survey," *IEEE Trans. Intell. Transport. Syst.*, vol. 15, no. 5, pp. 1869-1885, 2014.

[47] P. Zhang, F. Yan, and C. Du, "A comprehensive analysis of energy management strategies for hybrid electric vehicles based on bibliometrics," *Renew Sustain Energy Rev*, vol. 48, pp. 88-104, 2015.

[48] C. Martinez, X. Hu, D. Cao, E. Velenis, B. Gao, and M. Wellers, M. "Energy management in plug-in hybrid electric vehicles: Recent progress and a connected vehicles perspective," *IEEE Trans. Veh. Technol.*, vol. 66, no. 6, pp. 4534-4549, 2017.

[49] M. Sabri, K. Danapalasingam, and M. Rahmat, "A review on hybrid electric vehicles architecture and energy management strategies," *Renew Sustain Energy Rev*, vol. 53, pp. 1433-1442, 2016.

[50] W. Enang, and C. Bannister, "Modelling and control of hybrid electric vehicles (A comprehensive review)," *Renew Sustain Energy Rev*, vol. 74, pp. 1210-1239, 2017.

[51] C. Samanta, S. Padhy, S. Panigrahi, and B. Panigrahi, "Hybrid swarm intelligence methods for energy management in hybrid electric vehicles," *IET Electr Syst Transp*, vol. 3, no. 1, pp. 22-29, 2013.

[52] C. Sun, F. Sun, and H. He, "Investigating adaptive-ECMS with velocity forecast ability for hybrid electric vehicles," *Applied energy*, vol. 185, pp. 1644-1653, 2017.

[53] C. Musardo, G. Rizzoni, Y. Guezennec, and B. Staccia, "A-ECMS: An adaptive algorithm for hybrid electric vehicle energy management," *European Journal of Control*, vol. 11, no. 4-5, pp. 509-524, 2005.

[54] J. Liu, and H. Peng, "Modeling and control of a power-split hybrid vehicle," *IEEE Trans. Contr. Syst. Technol.*, vol. 16, no. 6, pp. 1242-1251, 2008.

[55] M. O'Keefe, and T. Markel, "Dynamic programming applied to investigate energy management strategies for a plug-in HEV," (No. NREL/CP-540-40376). National Renewable Energy Lab.(NREL), Golden, CO (United States), 2006.

[56] B. Chen, Y. Wu, and H. Tsai, "Design and analysis of power management strategy for range extended electric vehicle using dynamic programming," *Applied Energy*, vol. 113, pp. 1764-1774, 2014.

[57] C. Xu, A. Al-Mamun, S. Geyer, and H. Fathy, "A dynamic programming-based real-time predictive optimal gear shift strategy for conventional heavy-duty vehicles," in *Proc. Annual American Control Conference* (ACC), pp. 5528-5535 June 2018.

[58] J. Fu, S. Song, Z. Fu, and J. Ma, "Real-time implementation of optimal control considering gear shifting and engine starting for parallel hybrid electric vehicle based on dynamic programming," *Optimal Control Applications and Methods*, vol. 39, no. 2, pp. 757-773, 2018.

[59] T. Liu, H. Yu, H. Guo, Y. Qin, and Y. Zou, "Online energy management for multimode plug-in hybrid electric vehicles," *IEEE Trans Ind. Informat.*, DOI: 10.1109/ TII.2018. 2880897.

[60] M. Marzband, F. Azarinejadian, M. Savaghebi, and J. Guerrero, "An optimal energy management system for islanded microgrids based on multiperiod artificial bee colony combined with Markov chain," *IEEE Systems Journal*, vol. 11, no. 3, pp. 1712-1722, 2017.

[61] T. Sousa, H. Morais, Z. Vale, P. Faria, and J. Soares, "Intelligent energy resource management considering vehicle-to-grid: A simulated annealing approach," *IEEE Trans. on Smart Grid*, vol. 3, no. 1, pp. 535-542, 2012.

[62] T. Nüesch, P. Elbert, M. Flankl, C. Onder, and L. Guzzella, "Convex optimization for the energy management of hybrid electric vehicles considering engine start and gearshift costs," *Energies*, vol. 7, no. 2, pp. 834-856, 2014.

[63] X. Hu, N. Murgovski, L. Johannesson, and B. Egardt, "Optimal dimensioning and power management of a fuel cell/battery hybrid bus via convex programming, *IEEE/ASME Trans. Mechatronics*, vol. 20, no. 1, pp. 457-468, 2015.

[64] B. Gao, W. Zhang, Y. Tang, M. Hu, M. Zhu, and H. Zhan, "Game-theoretic energy management for residential users with dischargeable plug-in electric vehicles," Energies, vol. 7, no. 11, pp. 7499-7518, 2014.

[65] C. Dextreit, F. Assadian, I. Kolmanovsky, J. Mahtani, and K. Burnham, K. "Hybrid electric vehicle energy management using game theory," No. 2008-01-1317, *SAE Technical Paper*, 2008.

[66] F. Syed, M. Kuang, M. Smith, S. Okubo, and H. Ying, "Fuzzy gain-scheduling proportional–integral control for improving engine power and speed behavior in a hybrid electric vehicle," *IEEE Trans. Veh. Technol.*,

vol. 58, no. 1, pp. 69-84, 2009.

[67] Burch S, Cuddy M, Markel T. ADVISOR 2.1 documentation. Nat Renew Lab 1999.

[68] X. Tang, D. Zhang, T. Liu, A. Khajepour, H. Yu, and H. Wang, "Research on the energy control of a dual-motor hybrid vehicle during engine start-stop process," *Energy*, vol. 166, pp. 1181-1193, 2019.

[69] X. Zeng, and J. Wang, "A parallel hybrid electric vehicle energy management strategy using stochastic model predictive control with road grade preview," *IEEE Trans. Contr. Syst. Technol.*, vol. 23, no. 6, pp. 2416-2423, 2015.

[70] H. Borhan, A. Vahidi, A. M. Phillips, M. L. Kuang, I. V. Kolmanovsky, and S. Di Cairano, "MPC-based energy management of a power-split hybrid electric vehicle", *IEEE Trans. Control Syst. Technol.*, vol. 20, no. 3, pp. 593-603, May 2012.

[71] B. Sampathnarayanan, L. Serrao, S. Onori, G. Rizzoni, and S. Yurkovich, "Model predictive control as an energy management strategy for hybrid electric vehicles," *In Proc. ASME 2009 Dynamic Systems and Control Conference*, pp. 249-256, American Society of Mechanical Engineers.

[72] Z. Song, J. Hou, and H. Hofmann, et al., "Sliding-mode and Lyapunov function-based control for battery/supercapacitor hybrid energy storage system used in electric vehicles," *Energy*, vol. 122, pp. 601-612, 2017.

[73] H. Tian, and Z. Lu, et al., "A length ratio based neural network energy management strategy for online control of plug-in hybrid electric city bus," *Applied Energy*, vol. 177, pp. 71-80, 2016.

[74] T. Liu, Y. Zou, D. Liu, and F. Sun, "Reinforcement learning of adaptive energy management with transition probability for a hybrid electric tracked vehicle," *IEEE Trans. Ind. Electron.*, vol. 62, no. 12, 7837-7846, 2015.

[75] T. Liu, Y. Zou, D. Liu, and F. Sun, F. "Reinforcement learning–based energy management strategy for a hybrid electric tracked vehicle," Energies, vol. 8, no. 7, pp. 7243-7260, 2015.

[76] Y. Zou, T. Liu, D. Liu, and F. Sun, "Reinforcement learning-based real-time energy management for a hybrid tracked vehicle," *Applied energy*, vol. 171, pp. 372-382, 2016.

[77] T. Liu, X. Hu, S. E. Li, and D. Cao, "Reinforcement learning optimized look-ahead energy management of a parallel hybrid electric vehicle," *IEEE/ASME Trans. Mechatronics*, vol. 22, no. 4, 1497-1507, 2017.

[78] T. Liu, and X. Hu, "A Bi-Level Control for Energy Efficiency Improvement of a Hybrid Tracked Vehicle," *IEEE Transactions on Industrial Informatics*, vol. 14, no. 4, pp. 1616-1625, 2018.

[79] X. Qi, Y. Luo, G. Wu, K. Boriboonsomsin, and M. Barth, "Deep reinforcement learning-based vehicle energy efficiency autonomous learning system," *in Proc. 2017 IEEE Intelligent Vehicles Symposium* (IV), Redondo Beach, CA, USA, June 11-14, 2017.

[80] Y. Hu, W. Li, K. Xu, T. Zahid, F. Qin, and C. Li, "Energy management strategy for a hybrid electric vehicle based on deep reinforcement learning," *Applied Sciences*, vol. 8, no. 2, pp. 187, 2018.

[81] Y. Li, H. He, J. Peng, and H. Zhang, "Power management for a plug-in hybrid electric vehicle based on reinforcement learning with continuous state and action spaces," *in Proc. 9th Interna-tional Conference on Applied Energy*, ICAE2017, Cardiff, UK, August 21-24, 2017.

[82] T. Liu, X. Hu, W. Hu, Y. Zou, "A heuristic planning reinforcement learning-based energy management for power-split plug-in hybrid electric vehicles," *IEEE Trans Ind. Informat.*, DOI: 10.1109/TII.2019.2903098.

[83] P. Zhao, Y. Wang, N. Chang, Q. Zhu, and X. Lin, "A deep reinforcement learning framework for optimizing fuel economy of hybrid electric vehicles," *in Proc. 23rd Asia and South Pacific Design Automation Conference* (ASP-DAC), Jan 22-25, 2018.

[84] R. Liessner, C. Schroer, A. Dietermann, and B. Baker, "Deep reinforcement learning for advanced energy

management of hybrid electric vehicles," *in Proc. the 10th International Conference on Agents and Artificial Intelligence* (ICAART 2018), vol. 2, pp. 61-72, 2018.

[85] Y. Hay, M. Kuang, and R. McGee, "Trip-oriented energy management control strategy for plug-in hybrid electric vehicles," *IEEE Trans. Control Syst. Technol.*, vol. 22, pp. 1323-1336, 2014.

[86] E. Ozatay, S. Onori, J. Wollaeger, U. Ozguner, G. Rizzoni, D. Filev, and S. Di Cairano, "Cloud-based velocity profile optimization for everyday driving: A dynamic-programming-based solution," *IEEE Trans. Intell. Transp. Syst.*, vol. 15, pp. 2491-2505, 2014.

[87] N.P. Reddy, D. Pasdeloup, M.K. Zadeh and R. Skjetne, "An intelligent power and energy management system for fuel cell/battery hybrid electric vehicle using reinforcement learning," 2019 *IEEE Transportation Electrification Conference and Expo* (ITEC), *IEEE*, pp. 1-6, 2019.

[88] R.C. Hsu, S. Chen, W. Chen and C. Liu, "A reinforcement learning based dynamic power management for fuel cell hybrid electric vehicle," 2016 *Joint 8th International Conference on Soft Computing and Intelligent Systems* (SCIS) and *17th International Symposium on Advanced Intelligent Systems* (ISIS), *IEEE*, pp. 460-464, 2016.

[89] H. Sun, Z. Fu, F. Tao, L. Zhu and P. Si, "Data-driven reinforcement-learning-based hierarchical energy management strategy for fuel cell/battery/ultracapacitor hybrid electric vehicles," *Journal of Power Sources*, vol. 455, 2020.

[90] J. Yuan, L. Yang and Q. Chen, "Intelligent energy management strategy based on hierarchical approximate global optimization for plug-in fuel cell hybrid electric vehicles," *International Journal of Hydrogen Energy*, vol. 43, pp. 8063-8078, 2018.

[91] T. Gole, A. Hange, R. Dhar, A. Bhurke and F. Kazi, "Reinforcement learning based energy management in hybrid electric vehicle," 2019 *International Conference on Power Electronics, Control and Automation* (ICPECA), *IEEE*, pp. 1-5, 2019.

[92] S.A. Kouche-Biyouki, S.M.A. Naseri-Javareshk, A. Noori and F. Javadi-Hassanehgheh, "Power management strategy of hybrid vehicles using sarsa method," *26th Iranian Conference on Electrical Engineering*, New York, pp. 946-950, 2018.

[93] X. Lin, B. Zhou and Y. Xia, "Online recursive power management strategy based on the reinforcement learning algorithm with cosine similarity and a forgetting factor," *IEEE Transactions on Industrial Electronics*, DOI: 10.1109/TIE.2020.2988189.

[94] Z. Kong, Y. Zou and T. Liu, "Implementation of real-time energy management strategy based on reinforcement learning for hybrid electric vehicles and simulation validation," PLoS ONE, 12 (7): e0180491.

[95] G. Du, Y. Zou, X. Zhang, Z. Kong, J. Wu and D. He, "Intelligent energy management for hybrid electric tracked vehicles using online reinforcement learning," *Applied Energy*, vol. 251, 2019.

[96] H. Guo, S. Du, F. Zhao, Q. Cui and W. Ren, "Intelligent energy management for plug-in hybrid electric bus with limited state space," *Processes*, vol. 7, 2019.

[97] I. Sanusi, A. Mills, G. Konstantopoulos and T. Dodd, "Power management optimisation for hybrid electric systems using reinforcement learning and adaptive dynamic programming," *2019 American Control Conference* (ACC), pp. 2608-2613, 2019.

[98] B. Xu, D. Rathod, D. Zhang, A. Yebi, X. Zhang, X. Li and Z. Filipi, "Parametric study on reinforcement learning optimized energy management strategy for a hybrid electric vehicle," *Applied Energy*, vol. 259, 2020.

[99] Y. Yin, Y. Ran, L. Zhang, X. Pan and Y. Luo, "An energy management strategy for a super-mild hybrid electric vehicle based on a known model of reinforcement learning," J*ournal of Control Science and*

*Engineering*, vol. 2019, pp. 1-12, 2019.

[100] Q. Zhu and Q.Wang, "Real-time energy management controller design for a hybrid excavator using reinforcement learning," *Journal of Zhejiang University-SCIENCE A*, vol. 18, pp. 855-870, 2017.

[101] L. Fan, Y. Zhang, H. Dou and R. Zou, "Design of an integrated energy management strategy for a plug-in hybrid electric bus," *Journal of Power Sources*, vol. 448, 2020.

[102] B. Xu, X. Hu, X. Tang, X. Lin, H. Li, D. Rathod and Z. Filipi, "Ensemble reinforcement learning-based supervisory control of hybrid electric vehicle for fuel economy improvement," *IEEE Transactions on Transportation Electrification*, vol. 6, no. 2, pp. 717-727, 2020.

[103] T. Liu, G. Du, Y. Zou and D. Cao, "Fast learning-based control for energy management of hybrid electric vehicles," *IFAC-PapersOnLine*, vol. 51, no. 31, pp. 595-600, 2018.

[104] H. Liu, X. Li, W. Wang, L. Han and C. Xiang, "Markov velocity predictor and radial basis function neural network-based real-time energy management strategy for plug-in hybrid electric vehicles," *Energy*, vol. 152 pp. 427-444, 2018.

[105] S. Inuzuka, F. Xu, B. Zhang and T. Shen, "Reinforcement learning based on energy management strategy for HEVs," *2019 IEEE Vehicle Power and Propulsion Conference* (VPPC), pp. 1-6, 2019.

[106] H. Chaoui, H. Gualous, L. Boulon and S. Kelouwani, "Deep reinforcement learning energy management system for multiple battery based electric vehicles," *2018 IEEE Vehicle Power and Propulsion Conference* (VPPC), pp. 1-6, 2018.

[107] C. Alaoui, "Hybrid vehicle energy management using deep learning," *2019 International Conference on Intelligent Systems and Advanced Computing Sciences* (ISACS), pp. 1-5, 2019.

[108] C. Yang and C. Zhou, "An energy management strategy of hybrid electric vehicles based on deep reinforcement learning," *International Journal of Engineering and Advanced Research Technology* (IJEART), vol. 5, no. 12, 2019.

[109] P. Wang, Y. Li, S. Shekhar and W.F. Northrop, "A deep reinforcement learning framework for energy management of extended range electric delivery vehicles," *2019 IEEE Intelligent Vehicles Symposium* (IV), pp. 1837-1842, 2019.

[110] J. Wu, H. He, J. Peng, Y. Li and Z. Li, "Continuous reinforcement learning of energy management with deep Q network for a power split hybrid electric bus," *Applied Energy*, vol. 222, pp. 799-811, 2018.

[111] X. Han, H. He, J. Wu, J. Peng and Y. Li, "Energy management based on reinforcement learning with double deep Q-learning for a hybrid electric tracked vehicle," *Applied Energy*, vol. 254, 2019.

[112] X. Qi, Y. Luo, G. Wu, K. Boriboonsomsin and M. Barth, "Deep reinforcement learning enabled self-learning control for energy efficient driving," *Transportation Research Part C: Emerging Technologies*, vol. 99, pp. 67-81, 2019.

[113] R. Liessner, A. Lorenz, J. Schmitt, A.M. Dietermann and B. Baker, "Simultaneous electric powertrain hardware and energy management optimization of a hybrid electric vehicle using deep reinforcement learning and bayesian optimization," *2019 IEEE Vehicle Power and Propulsion Conference* (VPPC), pp. 1-6, 2019.

[114] P. Wang, Y. Li, S. Shekhar and W.F. Northrop, "Actor-Critic based deep reinforcement learning framework for energy management of extended range electric delivery vehicles," *2019 IEEE/ASME International Conference on Advanced Intelligent Mechatronics* (AIM), pp. 1379-1384, 2019.

[115] Y. Li, H. He, J. Peng and H. Wang, "Deep reinforcement learning-based energy management for a series hybrid electric vehicle enabled by history cumulative trip information," *IEEE Transactions on Vehicular Technology*, vol. 68, pp. 7416-7430, 2019.

[116] Y. Li, H. He, A. Khajepour, H. Wang and J. Peng, "Energy management for a power-split hybrid electric

bus via deep reinforcement learning with terrain information," *Applied Energy*, vol. 255, 2019.

[117] B. Bäker, A. Dietermann, J. Schmitt and R. Liessner, "Hyperparameter optimization for deep reinforcement learning in vehicle energy management," *Proceedings of the 11th International Conference on Agents and Artificial Intelligence*, pp. 134-144, 2019.

[118] R. Lian, J. Peng, Y. Wu, H. Tan and H. Zhang, "Rule-interposing deep reinforcement learning based energy management strategy for power-split hybrid electric vehicle," *Energy*, vol. 197, 2020.

[119] R. Liessner, A.M. Dietermann and B. Bäker, "Safe deep reinforcement learning hybrid electric vehicle energy management," *Agents and Artificial Intelligence*, pp. 161-181, 2019.

[120] S. Onori, and L. Tribioli, "Adaptive Pontryagin's Minimum Principle supervisory controller design for the plug-in hybrid GM Chevrolet Volt," *Applied Energy*, vol. 147, pp. 224-234, 2015.

[121] S. Ho, Y. Wong, and V. Chang, "Developing Singapore Driving Cycle for passenger cars to estimate fuel consumption and vehicular emissions," *Atmospheric environment*, vol. 97, pp. 353-362, 2014.

[122] E. Tazelaar, J. Bruinsma, B. Veenhuizen, and P. van den Bosch, "Driving cycle characterization and generation, for design and control of fuel cell buses," *World Electric Vehicle Journal*, vol. 3, no. 4, pp. 812-819, 2009.

[123] A. Ravey, N. Watrin, B. Blunier, D. Bouquain, and A. Miraoui, "Energy-source-sizing methodology for hybrid fuel cell vehicles based on statistical description of driving cycles," *IEEE Trans. Veh. Technol.*, vol. 60, no. 9, pp. 4164-4174, 2011.

[124] A. Ivanco, A. Charlet, Y. Chamaillard, and P. Higelin, "Energy management strategies for hybrid-pneumatic engine studied on a Markov chain type generated driving cycle," No. 2009-01-0145, *SAE Technical Paper*, 2009.

[125] H. Gong, Y. Zou, Q. Yang, J. Fan, F. Sun, and D. Goehlich, "Generation of a driving cycle for battery electric vehicles: A case study of Beijing," *Energy*, vol. 150, pp. 901-912, 2018.

[126] E. Torp, and P. Önnegren, "Driving cycle generation using statistical analysis and markov chains," Department of Electrical Engineering, Linköpings universitet, Sweden, 2013.

[127] M. Perhinschi, C. Marlowe, S. Tamayo, J. Tu, and W. Wayne, "Evolutionary algorithm for vehicle driving cycle generation," *Journal of the Air & Waste Management Association*, vol. 61, no. 9, pp. 923-931, 2011.

[128] J. Wu, C. Zhang, and N. Cui, "Fuzzy energy management strategy for a hybrid electric vehicle based on driving cycle recognition," *International journal of automotive technology*, vol. 13, no. 7, pp. 1159-1167, 2012.

[129] P. Nyberg, E. Frisk, and L. Nielsen, "Generation of equivalent driving cycles using Markov chains and mean tractive force components," *IFAC Proceedings Volumes*, vol. 47, no. 3, pp. 8787-8792, 2014.

[130] P. Nyberg, E. Frisk, and L. Nielsen, "Using real-world driving databases to generate driving cycles with equivalence properties," *IEEE Trans. Veh. Technol.*, vol. 65, no. 6, pp. 4095-4105, 2016.

[131] D. Huang, H. Xie, H. Ma, and Q. Sun, "Driving cycle prediction model based on bus route features," *Transportation Research Part D: Transport and Environment*, vol. 54, pp. 99-113, 2017.

[132] Y. Wang, W. Wang, Y. Zhao, L. Yang, and W. Chen, "A fuzzy-logic power management strategy based on Markov random prediction for hybrid energy storage systems," *Energies*, vol. 9, no. 1, 25, 2016.

[133] J. Oliva, C. Weihrauch, and T. Bertram, "Model-based remaining driving range prediction in electric vehicles by using particle filtering and Markov chains," *In Proc. 2013 World Electric Vehicle Symposium and Exhibition* (EVS27), pp. 1-10, 2013.

[134] R. Wang, and S. Lukic, "Review of driving conditions prediction and driving style recognition based control algorithms for hybrid electric vehicles," *In Proc. 2011 IEEE Vehicle Power and Propulsion*

*Conference*, pp. 1-7, 2011.

[135]  F. Bender, M. Kaszynski, and O. Sawodny, "Drive cycle prediction and energy management optimization for hybrid hydraulic vehicles," *IEEE Trans. Veh. Technol.*, vol. 62, no. 8, pp. 3581-3592, 2013.

[136]  Z. Chen, L. Li, B. Yan, C. Yang, C. Martínez, and D. Cao, "Multimode energy management for plug-in hybrid electric buses based on driving cycles prediction," *IEEE Trans. Intell. Transport. Syst.*, vol. 17, no. 10, pp. 2811-2821, 2016.

[137]  F. Tianheng, Y. Lin, G. Qing, H. Yanqing, Y. Ting, and Y. Bin, "A supervisory control strategy for plug-in hybrid electric vehicles based on energy demand prediction and route preview," *IEEE Trans. Veh. Technol.*, vol. 64, no. 5, pp. 1691-1700, 2015.

[138]  J. Valera, B. Heriz, G. Lux, J. Caus, and B. Bader, "Driving cycle and road grade on-board predictions for the optimal energy management in EV-PHEVs," *In Proc. 2013 World Electric Vehicle Symposium and Exhibition* (EVS27), pp. 1-10, 2013.

[139]  C. Xiang, F. Ding, W. Wang, and W. He, "Energy management of a dual-mode power-split hybrid electric vehicle based on velocity prediction and nonlinear model predictive control," *Applied energy*, vol. 189, pp. 640-653, 2017.

[140]  L. Niu, H. Yang, and Y. Zhang, "Intelligent HEV fuzzy logic control strategy based on identification and prediction of drive cycle and driving trend," *World Journal of Engineering and Technology*, vol. 3, no. 3, 215, 2015.

[141]  Z. Chen, R. Xiong, C. Wang, and J. Cao, "An on-line predictive energy management strategy for plug-in hybrid electric vehicles to counter the uncertain prediction of the driving cycle," *Applied energy*, vol. 185, pp. 1663-1672, 2017.

[142]  M. Zulkefli, J. Zheng, Z. Sun, and H. Liu, "Hybrid powertrain optimization with trajectory prediction based on inter vehicle communication and vehicle infrastructure integration," *Transportation Research Part C: Emerging Technologies*, vol. 45, pp. 41-63, 2014.

[143]  M. Debert, G. Yhamaillard, and G. Ketfi-herifellicaud, "Predictive energy management for hybrid electric vehicles-prediction horizon and battery capacity sensitivity," *IFAC Proceedings Volumes*, vol. 43, no. 7, pp. 270-275, 2010.

[144]  M. Andre, A. Hickman, D. Hassel, and R. Joumard, "Driving cycles for emission measurements under European conditions," *SAE transactions*, 562-574, 1995.

[145]  R. Nicolas, "The different driving cycles," Car Engineer, 2013: www.car-engineer.com/ the-different-driving-cycles/.

[146]  E. Tzirakis, K. Pitsas, F. Zannikos, and S. Stournas, "Vehicle emissions and driving cycles: comparison of the Athens driving cycle (ADC) with ECE-15 and European driving cycle (EDC)," *Global NEST Journal*, vol. 8, no. 3. pp. 282-290, 2006.

[147]  M. André, "The ARTEMIS European driving cycles for measuring car pollutant emissions," *Science of the total Environment*, vol. 334, pp. 73-84, 2004.

[148]  "Transportation Secure Data Center," National Renewable Energy Laboratory (NREL). Accessed Jan. 15, 2019: www.nrel.gov/tsdc, 2019.

[149]  E. Burton, J. Gonder, A. Duran, and E. wood, "Map matching and real world integrated sensor data warehousing", 2013 Federal Committee on Statistical Methodology (FCSM) Research Conference, November 4-6, Washington, DC, 2013.

[150]  K. Regimbal, I. Carpenter, C. Chang, and S. Hammond, "Peregrine at the national renewable energy laboratory," *Contemporary High Performance Computing: From Petascale toward Exascale*, Vol. 2, no. 23,

163, 2015.

[151] B. Coifman, and L. Li, "A critical evaluation of the Next Generation Simulation (NGSIM) vehicle trajectory dataset," *Transportation Research Part B: Methodological*, vol. 105, pp. 362-377, 2017.

[152] L. Zhang, V. Kovvali, N. Clark, D. Sallman, and V. Alexiadis, "NG-Video user's manual," FHWA-HOP-07-009, Oct. 2006.

[153] X. Lu, and A. Skabardonis, "Freeway traffic shockwave analysis: exploring the NGSIM trajectory data," *In Proc. 86th Annual Meeting of the Transportation Research Board*, Washington, DC, Jan. 2007.

[154] M. Montanino, and V. Punzo, "Making NGSIM data usable for studies on traffic flow theory: Multistep method for vehicle trajectory reconstruction," Transportation Research Record, vol. 2390, no. 1, pp. 99-111, 2013.

[155] R. Krajewski, J. Bock, L. Kloeker, and L. Eckstein, "The highd dataset: A drone dataset of naturalistic vehicle trajectories on german highways for validation of highly automated driving systems," *In Proc. 2018 21st International Conference on Intelligent Transportation Systems* (ITSC), pp. 2118-2125, Nov. 2018.